\documentclass[11pt,letterpaper]{article}

\pdfoutput=1

\usepackage[sortcites=false]{newchet}
\defbibheading{bibliographyintoc}[\refname]{\printbibheading[heading=bibliography,title={#1}]\phantomsection\addcontentsline{toc}{section}{#1}}
\allowdisplaybreaks[0]

\renewcommand{\acknowledgementsname}{Acknowledgments}

\usepackage{dsfont}
\usepackage{mathrsfs}
\usepackage{tikz}
\usepackage{tabularx}
\usepackage{braket}
\usepackage{tensor}
\usepackage{mathtools}

\renewcommand\braket[1]{{\langle{#1}\rangle}}

\usetikzlibrary{intersections,calc,arrows.meta,bending,patterns,decorations.pathmorphing,decorations.markings,colorbrewer}

\hypersetup{pdfauthor={Thomas Hartman, Gr\'egoire Mathys and Andreas Stergiou}}

\newcommand{\traceT}{T}
\newcommand{\opTheta}{\Theta}

\newcommand{\opS}{\mathsf{S}}

\newcommand{\lsp}{\hspace{0.5pt}}
\newcommand{\logsp}{\hspace{-0.5pt}}

\renewcommand{\geq}{\geqslant}
\renewcommand{\leq}{\leqslant}
\renewcommand{\opS}{\mathcal{S}}

\def\gsim{ \lower .75ex \hbox{$\sim$} \llap{\raise .27ex \hbox{$>$}} }
\def\lsim{ \lower .75ex \hbox{$\sim$} \llap{\raise .27ex \hbox{$<$}} }
\def\be{\begin{equation}}
\def\ee{\end{equation}}
\def\bea{\begin{eqnarray}}
\def\eea{\end{eqnarray}}

\newcommand{\baaa}{\begin{eqnarray}}
\newcommand{\eaaa}{\end{eqnarray}}

\DeclareMathOperator{\E}{e}

\newcommand{\p}{\partial}
\renewcommand{\O}{\mathcal{O}}
\renewcommand{\E}{\mathcal{E}}

\DeclareSymbolFont{matha}{OML}{txmi}{m}{it}
\DeclareMathSymbol{v}{\mathord}{matha}{118}

\renewcommand{\O}{{\mathcal O}}

\newcommand{\rhoE}{\rho^{\cal E}}
\newcommand{\GE}{G^{\cal E}}
\newcommand{\FE}{F^{\cal E}}
\newcommand{\D}{\mu\frac{d}{d\mu}}
 
\DeclareMathOperator{\Disc}{Disc}
\DeclareMathOperator{\dDisc}{dDisc}

\makeatletter
\newsavebox{\@brx}
\newcommand{\llangle}[1][]{\savebox{\@brx}{\(\m@th{#1\langle}\)}%
\mathopen{\copy\@brx\kern-0.5\wd\@brx\usebox{\@brx}}}
\newcommand{\rrangle}[1][]{\savebox{\@brx}{\(\m@th{#1\rangle}\)}%
\mathclose{\copy\@brx\kern-0.5\wd\@brx\usebox{\@brx}}}
\renewcommand\@makefntext[1]{\parindent 1em\noindent\@makefnmark#1}
\makeatother

\preprint{CERN-TH-2026-288}

\title{A Positive Metric on Four-Dimensional QFTs}

\author{Thomas Hartman,${\!}^a$ Gr\'egoire Mathys,${\!}^{b,c}$ and Andreas Stergiou$^d$}

\affiliation{$^a$Leinweber Institute for Theoretical Physics, Stanford University, Stanford, CA 94305, USA\\[3pt]
$^b$Theoretical Physics Department, CERN, CH-1211 Geneva 23, Switzerland\\[3pt]
$^c$Fields and Strings Laboratory, Institute of Physics,\\[-3pt]
\'Ecole Polytechnique F\'ed\'erale de Lausanne (EPFL), CH-1015 Lausanne, Switzerland\\[3pt]
$^d$Department of Mathematics, King's College London, Strand, London WC2R 2LS, United Kingdom}

\email{tomhartman@stanford.edu, gregoire.mathys@cern.ch, andreas.stergiou@kcl.ac.uk}

\date{November 2026}

\abstract{A natural metric on the space of four-dimensional quantum field theories was introduced by Jack and Osborn in classic work on the local renormalization group. In perturbation theory, the Jack--Osborn metric determines the flow of the $a$-anomaly, and positivity implies the $a$-theorem. We derive a non-perturbative generalization of the Jack--Osborn metric and show that it is non-negative by the averaged null energy condition (ANEC). Consequently, the ANEC implies the strong $a$-theorem, and provides the monotonic function along the flow. The method is based on spectral analysis of the three-point function $\langle \O_I \E \O_J\rangle$, where the $\O_I$ are relevant operators and $\E$ is the averaged null energy. This three-point function is decomposed under the spectrum of null rotations that fix the light ray, leading to a positivity condition at each impact parameter.}

\begin{document}

\maketitle

\toc

\section{Introduction}

Consider a space of unitary QFTs, $S_0 + \lambda^I \int d^{\lsp d} x\,\O_I$, where the operators $\O_I$ are marginal and the couplings $\lambda^I$ provide local coordinates. The Zamolodchikov metric $G_{IJ} = \langle \O_I(0) \O_J(1)\rangle$ is one natural metric on this space. In two dimensions, it controls the RG flow of the $c$-function by the relation \cite{Zamolodchikov:1986gt,Cappelli:1990yc}
\begin{equation}\label{introcflow}
    \mu \frac{d}{d\mu}\tilde{c} = 6\pi^2 G_{IJ} \beta^I \beta^J\,,
\end{equation}
where $\mu$ is the renormalization scale. The Zamolodchikov metric is manifestly positive-definite by reflection positivity, so the $c$-function is monotonic. This is the proof of the strong $c$-theorem in two dimensions. The `strong' version of the $c$-theorem states that there is a monotonic function along the flow that agrees with the central charge at the fixed points, and the `weak' version is the inequality $c_{\text{UV}} > c_{\text{IR}}$.

In four dimensions, there is an analogous flow equation for a function $\tilde{a}(\lambda)$ that is equal to the $a$-anomaly at fixed points. The flow equation takes the same form \cite{Jack:1990eb},
\begin{equation}\label{introaflow}
    \mu \frac{d}{d\mu} \tilde{a} = H_{IJ} \beta^{I} \beta^J\,.
\end{equation}
However, $H_{IJ}$ is not the Zamolodchikov metric, but a different object that we refer to as the Jack--Osborn metric. $H_{IJ}$ is  not manifestly positive. In their original work on the subject, Jack and Osborn found that $H_{IJ} \propto G_{IJ} + O(\beta, \p \beta)$, which establishes a perturbative version of the strong $a$-theorem. The non-perturbative $a$-theorem, first proposed by Cardy \cite{Cardy:1988cwa}, has now been proved several different ways: First using dilaton scattering \cite{Komargodski:2011vj}, then by entanglement inequalities \cite{Casini:2017vbe}, and more recently from the averaged null energy condition (ANEC) \cite{Hartman:2023qdn,Hartman:2024xkw}. The derivation based on the ANEC in previous work gave the weak $a$-theorem, but no monotonic $a$-function.

In this paper, we derive a non-perturbative generalization of the Jack--Osborn metric in four dimensions.  The main idea is that the Zamolodchikov metric $\langle \O_I \O_J\rangle$ in 2d is replaced by a three-point function $\langle \O_I \E \O_J\rangle$ in 4d. Here $\E$ is the ANEC operator through the origin,
\begin{align}
    \E = \int du\, T_{uu} \, ,
\end{align}
where $u$ is a null direction. 
Unlike the 2-point function, the 3-point function has non-trivial kinematics, and we cannot simply evaluate $\langle \O_I \E \O_J\rangle$ at fixed positions. Instead, we will find a natural spectral decomposition of this three-point function and define the metric from the spectral function. The result is a fully non-perturbative metric that is non-negative by the ANEC. This also upgrades the ANEC sum rule derived in \cite{Hartman:2023qdn} from the weak $a$-theorem to the strong $a$-theorem.

The spectral decomposition of the three-point function $\langle \O_I \E \O_J\rangle$ starts from the observation that the ANEC operator $\E$ commutes with a set of Poincar\'e charges known as null rotations. An ordinary rotation fixes an axis; a null rotation fixes a null ray. The null rotation charge is closely related to impact parameter, and thus we obtain a positive spectral representation of the 3-point function in impact parameter space. There is a close correspondence between the Jack--Osborn flow equation in 4d, written in terms of the impact parameter, and the Zamolodchikov flow equation in 2d. See Table \ref{tab:comparison}.

We will then revisit the local RG formalism in the context of perturbative QFTs. In a CFT in even spacetime dimensions, the trace of the stress tensor in a general background metric $g_{\mu\nu}$ is $\langle T \rangle_{g} = {\cal A}(g)$, where ${\cal A}$ is the Weyl anomaly. Needless to say, the fact that ${\cal A}$ is a local diffeomorphism-invariant function of background fields is very powerful. The starting point of the local RG technique is to recognize that in a non-conformal but renormalizable QFT, there is a similar equation that is almost as powerful:
\begin{equation}
    \langle T - \beta^I \O_I \rangle_{g,\lambda} = {\cal A}(g,\lambda)\,.
\end{equation}
The couplings have been promoted to spacetime-dependent background fields $\lambda^I(x)$, and once again ${\cal A}$ is local. The different terms appearing in ${\cal A}$ are related by Wess--Zumino consistency conditions~\cite{Osborn:1991gm}, and one such condition is the flow equation \eqref{introaflow}. This extends to any even $d$ \cite{Grinstein:2013cka}, but positivity of the associated metric has been shown only perturbatively around free theories, particularly in $d=4$ \cite{Jack:1990eb,Baume:2014rla} and $d=6$ \cite{Grinstein:2014xba,Grinstein:2015ina,Stergiou:2016uqq}. For comprehensive discussions of the local RG and Weyl consistency conditions see \cite{Jack:2013sha,Baume:2014rla}, and for a review see \cite{Shore:2016xor}.

We will show that the local RG equations in $d=4$ simplify dramatically when applied to the averaged null energy operator $\E$. We will then use these results to study the positivity of the Jack--Osborn metric, $H_{IJ}$. The metric is scheme-dependent, so the question is whether there exists a scheme in which $H_{IJ}$ is positive. We find a scheme where $H_{IJ}$ is equal to the non-perturbative ANEC spectral function, thereby establishing positivity.

The connection between the perturbative Jack--Osborn metric and dilaton scattering amplitudes was studied in \cite{Baume:2014rla}. There is a scheme in which the Jack--Osborn metric is a positive unitarity cut of a forward amplitude, and in this scheme, the monotonic $a$-function of Jack and Osborn agrees perturbatively with that of Komargodski and Schwimmer \cite{Komargodski:2011vj} (see also \cite{Karateev:2023mrb,Karateev:2024skm}). Our analysis of the local RG provides analogous results for the ANEC sum rule. 

One motivation for this work is to set the stage for studying global properties of the space of QFTs. See \cite{Vafa:1988ue,Kutasov:1988xb,Friedan:2009ik,Douglas:2010ic,Friedan:2012hi,Gukov:2015qea,Balthazar:2022hzb} for results in two dimensions, and \cite{Ooguri:2006in,Heidenreich:2018kpg,Perlmutter:2020buo,Baume:2023msm,Ooguri:2024ofs} for possible ties to the swampland program and the asymptotic limits of theory space --- in four dimensions, the asymptotic behavior is typically measured in the Zamolodchikov metric, but could the Jack--Osborn metric be relevant? 

More broadly, there is an intriguing web of connections among $c$-theorems, quantum information, and energy conditions. There are entropic proofs of $c$-theorems in two, three, and four dimensions \cite{Casini:2004bw,Casini:2012ei,Casini:2017vbe,Casini:2023kyj,Kudler-Flam:2026abf}; an information-theoretic proof of the ANEC \cite{Faulkner:2016mzt}, which also follows from causality \cite{Hartman:2015lfa,Hartman:2016dxc,Hartman:2016lgu}; holographic results that tie them together \cite{Erdmenger:2001ja,Myers:2010xs,Kelly:2014mra}; direct relations between the Zamolodchikov metric and an information metric \cite{Balasubramanian:2014bfa,Stout:2021ubb,Stout:2022phm}; and various extensions to defects and other settings \cite{Jensen:2015swa,Balakrishnan:2017bjg,Belin:2020lsr,Besken:2020snx,Casini:2023kyj,Meineri:2023mps,Nakamura:2025hyw}. Light-ray operators such as the ANEC provide constraints on CFTs and control their Regge behavior \cite{Hofman:2008ar,Hofman:2009ug,Hofman:2016awc,Cordova:2017zej,Afkhami-Jeddi:2017rmx,Kravchuk:2018htv,Belin:2019mnx,Kologlu:2019bco,Kologlu:2019mfz,Mecaj:2025ecl,Belin:2026wkc}. They have also seen a revival in particle physics, where light-ray operators model calorimeters, and provide insight into recent collider experiments \cite{Lee:2022ige,Moult:2025nhu}. The parallels to two dimensions indicated in Table \ref{tab:comparison} are also reminiscent of celestial holography, where light ray operators naturally organize asymptotic symmetries and appear in soft theorems \cite{Cordova:2018ygx,Hu:2022txx,Hu:2023geb,Himwich:2025ekg,Himwich:2026exq,Strominger:2026yrh}.

A natural question is whether our methods extend to higher dimensions. There is evidence for a six-dimensional $a$-theorem \cite{Cordova:2015fha,Stergiou:2016uqq}, and to a lesser extent a five-dimensional $F$-theorem \cite{Giombi:2014xxa,Myers:2010tj}. The quantity $\Delta a $ in six dimensions can be extracted by a light-ray sum rule, but we have not found a way to prove positivity. There are analogous, inconclusive results from dilaton amplitudes \cite{Elvang:2012st}.

\subsection{Outline and summary of results} 

\begin{table}[t]
\centering
\captionsetup{format=plain}
\renewcommand{\arraystretch}{1.6}
\newcolumntype{C}{>{\centering\arraybackslash}X}
\begin{tabularx}{\textwidth}{@{}>{\sffamily\bfseries\small}l C C@{}}
\toprule
                   & $d=2$ & $d=4$ \\
\midrule
correlator         & $\langle \O_I(x)\O_J(0)\rangle $ &  $\langle \O_I(x)\mathcal{E}\O_J(y)\rangle $ \\
positivity         & reflection positivity & ANEC  \\
scale              & separation $|\vec{x}|$ & impact parameter $|\vec{b}|$ \\
sum rule           & $\displaystyle \Delta c= 3\pi \int d^2\vec{x}\, \vec{x}^2\langle T(x)T(0)\rangle $& $\displaystyle \Delta a= \frac{1}{32} \int d^2\vec{b}\, \vec{b}\,^2\rho_{TT}^\mathcal{E}(b) $ \\
monotonic function & integrate over $|\vec{x}| > \mu^{-1}$ & integrate over $|\vec{b}|>\mu^{-1}$ \\
scheme             & $\displaystyle \int_{|x|<\mu^{-1}} d^2\vec{x}\, \vec{x}^2 \langle \O_I(x)\O_J(0)\rangle =0 $ &  $\displaystyle \int_{|\vec{b}|<\mu^{-1}} d^2\vec{b}\, \vec{b}^{\,2} \FE_{IJ}(b) =0 $\\
local RG metric    & $\displaystyle G_{IJ}=\frac{\pi}{2\mu^4}\langle \O_I(\mu^{-1})\O_J(0)\rangle $ & $\displaystyle H_{IJ} = \frac{\pi}{16\mu^4}\rho_{IJ}^{\mathcal{E}}(\mu^{-1})$ \\
\bottomrule
\end{tabularx}
\caption{
    \label{tab:comparison}
    Comparison between the $d=2$ and $d=4$ cases. By organizing the computation in terms of impact parameter $\vec{b}$, the four-dimensional case closely mimics the two-dimensional case. $\FE_{IJ}(b)$ is $\langle \O_I \E \O_J\rangle$ in impact parameter space, see \eqref{eq:impparcorrs} below.}
\end{table}

In section \ref{s:spectral} we study the spectral decomposition of three-point functions involving the ANEC, such as $\langle \O_I \E \O_J\rangle$. In $d$ dimensions, the ANEC operator $\E$ commutes with the $u$-momentum $P_u$ and with $d-2$ transverse null rotations, which are the generators of the Poincar\'e group that fix the null ray. The spectral function is defined by matrix elements of $\E$ on the subspace of definite null momentum. We will show that this is captured by a spectral function
\begin{align}
    \rhoE_{IJ}(b)  \succeq 0 \,,
\end{align}
where $\vec{b} \in \mathbb{R}^{d-2}$ is the impact parameter in the transverse space and $b=|\vec{b}|$. This is positive semidefinite as a matrix in the indices $I,J$ at any fixed $\vec{b}$. We also relate the spectral function to the time-ordered correlator $\langle {\cal T}[\O_I\O_J\E]\rangle$ and find that, in impact parameter space, they are equal up to a sign when $\vec{b} \neq 0$. 

In section \ref{s:exacta}, we apply these results to re-interpret the ANEC sum rule derived in \cite{Hartman:2023qdn}. We show that the sum rule can be written as
\begin{align}\label{introdeltaa}
    \Delta a= \frac{1}{32} \int d^2 \vec{b} \, \vec{b}^{\,2} \rhoE_{TT}(b) \, , 
\end{align}
where $\rhoE_{TT}$ is the spectral function for $\langle T \E T\rangle$. Not only is this positive (the result of \cite{Hartman:2023qdn}), but the integrand is positive for all $\vec{b}$. This allows us to write a monotonic $a$-function by cutting off the integral at $|\vec{b}| = \mu^{-1}$,
\begin{equation}\label{npafunctionintro}
    \tilde{a}(\mu) = a_{\text{IR}} + \frac{1}{32}\int_{|\vec{b}| > \mu^{-1}} d^{\lsp 2} \vec{b} \, \vec{b}^{\,2} \rhoE_{TT}(b)\,.
\end{equation}
 The metric inherited from this monotonic $a$-function is 
\begin{align}\label{posmet}
    H_{IJ} = \frac{\pi}{16\mu^4} \rhoE_{IJ}(\mu^{-1})\,.
\end{align}
This is the non-perturbative positive metric along an RG flow.
The indices $I,J$ run over the relevant operators (which in the non-perturbative setting do not need to have small anomalous dimensions). 

The ANEC sum rule in the form \eqref{introdeltaa} is closely related to, and directly inspired by, the recent paper by OpenAI on scale vs conformal invariance \cite{OpenAI_SvC}. It appears that Proposition 3.4 therein is essentially \eqref{introdeltaa} in disguise.

In sections \ref{s:localrgsetup}--\ref{s:jometric} we study perturbative RG flows using the local RG formalism. The formalism is introduced briefly in section \ref{s:localrgsetup}, followed in section \ref{s:localrg2d} by its application to 2d RG flows. Most of the discussion in two dimensions is review, as the main results already appear in \cite{Zamolodchikov:1986gt,Cappelli:1990yc,Osborn:1991gm}, but our presentation is different and more easily adapted to the study of the ANEC in four dimensions. In particular we describe how to choose a scheme in which the Jack--Osborn $c$-function becomes Zamolodchikov's monotonic $c$-function.

In section \ref{s:localrg4d} we study the local RG in four dimensions. We review the results of \cite{Osborn:1991gm} and rephrase them in terms of correlation functions. The full set of RG equations for 3-point functions is very complicated, but we find simple, compact expressions for correlators involving the ANEC operator. This section has two main outputs. First, we show that the flow equation \eqref{introaflow} can be re-interpreted as the Callan--Symanzik equation
\begin{align}
    \mu \frac{d}{d\mu} \langle T \E T \rangle =0 \, .
\end{align}
Second, we show how the Jack--Osborn $a$-function $\tilde{a}$ can be easily extracted from the 3-point functions involving $\E$. 

In section \ref{s:jometric} we unify the results of section \ref{s:exacta} on the non-perturbative, ANEC-derived $a$-function, and section \ref{s:localrg4d} on the perturbative $a$-function from the local RG. In the local RG, the Jack--Osborn metric and the $a$-function along the flow are scheme-dependent, so the question is not whether the metric is positive, but whether there exists a scheme in which it is positive. We find a renormalization scheme in which the 4d Jack--Osborn metric is given by \eqref{posmet} and is therefore non-negative. This is directly analogous to the situation in two dimensions, where there is a particular scheme in which the local RG metric equals the Zamolodchikov metric.

It is an open question whether the renormalization group in four dimensions is a gradient flow on the $a$-function landscape, which would be the strongest version of the strong $a$-theorem \cite{Friedan:2009ik,Barnes:2004jj}. We comment on the gradient property in section \ref{s:gradient}, but do not find a non-perturbative proof of the gradient property.

Conventions for correlation functions, in Euclidean and Lorentzian signature, can be found in appendix \ref{app:corr}. 
In appendix \ref{app:momentum}, we study the spectral decomposition in momentum space instead of  impact parameter space. The ANEC operator is non-diagonal in momenta, so the spectral decomposition has more parameters. The main result of appendix \ref{app:momentum}, which is not used elsewhere in this work, is a formula relating the momentum-space spectral kernel to a double discontinuity.

\section{Spectral analysis of the energy operator}\label{s:spectral}

In four-dimensional Minkowski space, introduce null coordinates $u=t-x$ and $v=t+x$ such that the metric is
\begin{equation}
    ds^2 = -du\, dv + d\vec{y}^{\,2}\,, \qquad \quad \vec{y} \in \mathbb{R}^2\,.
\end{equation}
We will use indices $x^\mu = (u,v,y^a)$ with $a=1,2$, and write momenta as
\begin{equation}
    k_\mu = (k_u, k_v, \vec{k})\,, 
\end{equation}
often using $k_u = -\omega$. The spectrum condition requires $-k_t=E\geq \sqrt{k_x^2+\vec{k}^{\lsp 2}}$, and $k_u=\frac12(k_t-k_x)$, which implies that physical states have $k_u \leq 0 $, i.e.\ $\omega\geq0$. The averaged null energy operator, or ANEC operator, is
\begin{equation}
    \E(v,\vec{y}) = \int_{-\infty}^{\infty}du\, T_{uu}\left(u,v,\vec{y}\right)\,,
\end{equation}
and the ANEC through the origin is denoted by the shorthand
\begin{equation}
    \E = \E(0,0)\,.
\end{equation}
The ANEC is non-negative and annihilates the vacuum: $\E|0\rangle = 0$ \cite{Hartman:2016lgu,Faulkner:2016mzt,Epstein:1965zza,Kravchuk:2018htv}. 

In this section we will study the spectral decomposition of 3-point functions involving $\E$, such as $\langle \O(x_1) \E \O(x_2)\rangle$. One approach is to project onto definite 4-momenta on each side of the ANEC operator. However, $\E$ does not commute with $P^\mu$. Instead, we will use an impact parameter representation with a maximal set of generators that commute with the ANEC. The momentum representation is studied in appendix \ref{app:momentum}.

\subsection{Light ray kinematics}

In this subsection only, we work in $d$ dimensions with $\vec{y} \in \mathbb{R}^{d-2}$. With the light ray placed through the origin, the centralizer of $\E$ within the Poincar\'e group  is generated by
\begin{equation}
    P_u\,, \quad J_{ab}\,, \quad\text{and}\quad N_{a}\,, \quad \text{ with } a,b=1,\dots,d-2\,.
\end{equation}
Here $(J_{ab}, N_a)$ generate the massless little group ISO$(d-2)$. $J_{ab}$ is a rotation in the transverse space, and the null rotation $N_a$ is the generator of
\begin{equation}
    \zeta_a = y_a \p_u + \frac{v}{2}\p_a\,. \label{eq:Killingzeta}
\end{equation}
Just as an ordinary rotation fixes an axis, a null rotation fixes a null ray. It can be expressed as a sum of a boost and a rotation; in the coordinates $(t,x,\vec{y})$, 
\begin{equation}
\zeta_a = \frac{1}{2}\Big[\underbrace{(y_a\partial_t + t\partial_a)}_{\text{boost along }y_a}+ \underbrace{(x\partial_a -y_a\partial_x)}_{\text{rotation in }(y_a,x)}\Big]\, .
\end{equation}
On the null plane $v=0$, $\zeta_a= y_a\partial_u$ as shown in Fig.~\ref{fig:verticalcell}. 
\begin{figure}[t]
    \centering
    \captionsetup{format=plain}
    \begin{minipage}[t]{0.485\linewidth}
        \centering
        \resizebox{\linewidth}{!}{
\begin{tikzpicture}[
  scale=1.2, transform shape,
  x={(-0.79cm,0.79cm)}, y={(0.73cm,0.20cm)}, z={(0.79cm,0.79cm)},
  font=\small, line cap=round, line join=round,
  >=Latex,
  nr axis/.style={draw=black!65,line width=0.65pt,-{Latex[length=2.2mm]}},
  nr before/.style={draw=black!45,line width=0.7pt,dash pattern=on 3pt off 2.5pt},
  nr after/.style={draw=nrblue,line width=1.05pt},
  nr rotation/.style={draw=nrblue,line width=0.7pt,-{Latex[length=1.7mm]}}
]
\definecolor{nrblue}{RGB}{38,103,137}
\definecolor{nranec}{RGB}{193,78,42}

\filldraw[fill=nrblue!7,draw=nrblue!36,line width=0.65pt]
  (-1.25,-1.8,0)--(3.8,-1.8,0)--(3.8,1.8,0)--(-1.25,1.8,0)--cycle;
\foreach \yy in {-1.5,-1,-0.5,0.5,1,1.5}
  \draw[nrblue!20,line width=0.4pt] (-1.25,\yy,0)--(3.8,\yy,0);

\def\nralpha{0.55}
\foreach \uu in {0.9,1.9,2.9} {
  \draw[nr before] (\uu,-1.5,0)--(\uu,1.5,0);
  \draw[nr after]
    ({\uu-1.5*\nralpha},-1.5,0)--({\uu+1.5*\nralpha},1.5,0);
  \foreach \yy in {-1.05,1.05}
    \draw[nr rotation]
      (\uu,\yy,0)--({\uu+\nralpha*\yy},\yy,0);
}

\draw[draw=black!40,line width=0.55pt] (-1.65,0,0)--(0,0,0);
\draw[nr axis] (0,0,0)--(4.55,0,0) node[above left=2pt] {$u$};
\draw[nr axis] (0,-1.62,0)--(0,2.35,0) node[above right=1pt] {$y$};
\draw[nr axis] (0,0,-0.42)--(0,0,3.16) node[above right=2pt] {$v$};

\draw[nranec,line width=1.8pt] (-1.25,0,0)--(3.8,0,0);
\node[above right=0pt,text=nranec] at (3.45,0,0) {$\E$};

\node[anchor=center,text=nrblue,font=\small] at (-0.55,-1.1,0)
  {$v=0$};
\end{tikzpicture}}
        \caption{A null rotation in $2+1$ dimensions. On the null plane $v=0$, the null rotation is $\zeta_a = y_a \p_u$. The null rotation charges $N_a$ and the momentum $P_u$ commute with each other and with the ANEC operator, $\E$.}
        \label{fig:verticalcell}
    \end{minipage}\hfill
    \begin{minipage}[t]{0.485\linewidth}
        \centering
        \resizebox{\linewidth}{!}{\begin{tikzpicture}[
  scale=1.2, transform shape,
  x={(-0.79cm,0.79cm)}, y={(0.73cm,0.20cm)}, z={(0.79cm,0.79cm)},
  font=\small, line cap=round, line join=round,
  >=Latex,
  nr axis/.style={
    draw=black!65,line width=0.65pt,
    -{Latex[length=2.2mm]}
  },
  nr particle/.style={
    draw=nrblue,line width=1.05pt
  },
  nr incoming/.style={
    draw=nrblue!75,line width=0.85pt,
    dash pattern=on 3pt off 2.5pt
  },
  nr impact/.style={
    draw=black!75,line width=0.7pt,
    {Latex[length=1.7mm]}-{Latex[length=1.7mm]}
  }
]
\definecolor{nrblue}{RGB}{38,103,137}
\definecolor{nranec}{RGB}{193,78,42}

\filldraw[fill=nrblue!7,draw=nrblue!36,line width=0.65pt]
  (-1.25,-1.8,0)--(3.8,-1.8,0)--
  (3.8,1.8,0)--(-1.25,1.8,0)--cycle;

\foreach \yy in {-1.5,-1,-0.5,0.5,1,1.5}
  \draw[nrblue!20,line width=0.4pt]
    (-1.25,\yy,0)--(3.8,\yy,0);

\def\nruc{1.70}
\def\nrb{1.40}
\coordinate (nrA) at (\nruc,0,0);
\coordinate (nrB) at (\nruc,\nrb,0);

\draw[nr incoming] (\nruc,\nrb,-2.40)--(nrB);

\draw[draw=black!40,line width=0.55pt]
  (-1.65,0,0)--(0,0,0);
\draw[nr axis]
  (0,0,0)--(4.55,0,0) node[above left=2pt] {$u$};
\draw[nr axis]
  (0,-1.62,0)--(0,2.35,0) node[above right=1pt] {$y$};
\draw[nr axis]
  (0,0,-0.42)--(0,0,3.16) node[above right=2pt] {$v$};

\draw[nranec,line width=1.8pt]
  (-1.25,0,0)--(3.8,0,0);
\node[above right=0pt,text=nranec]
  at (3.45,0,0) {$\E$};

\node[anchor=center,text=nrblue,font=\small]
  at (-0.55,-1.1,0) {$v=0$};

\draw[nr impact,shorten <=2pt,shorten >=2pt]
  (nrA)--(nrB)
  node[midway,above=3pt,text=black] {$b$};

\draw[nr particle,-{Latex[length=2.3mm]}]
  (nrB)--(\nruc,\nrb,2.15)
  node[above=4pt,text=nrblue] {$P_u=-\omega,\quad N=\omega b$};

\fill[nrblue] (nrB) circle[radius=1.8pt];

\end{tikzpicture}}
        \caption{Impact parameter in $2+1$ dimensions. A particle with momentum $P_u=-\omega$ crosses the null plane $v=0$ at a transverse distance $y=b$ from the ANEC operator $\E$. Its null rotation charge is $N=\omega b$.}
        \label{fig:impactparam}
    \end{minipage}
\end{figure}
The null rotation charges can be expressed in terms of the ANEC operator by evaluating them on the $v=0$ plane,
\begin{equation}
    N_a = \int d^{\lsp d-2} \vec{y} \, y_a \, \E(0,\vec{y})\,.
\end{equation}
If a particle with $P_u = -\omega$ crosses the null plane $v=0$ at impact parameter $\vec{y} = \vec{b}$, then it has $N_a= \omega b_a$, as illustrated in Fig.~\ref{fig:impactparam}. Therefore in states of fixed $\omega$, we can interpret the null rotation as the (dimensionless) impact parameter.

The operators $(\E, P_u, N_1, N_2,\dots, N_{d-2})$ mutually commute. The commutation relation of a null rotation with a scalar $\O$ is
\begin{equation}
    [N_a, \O(x)] = - i\left( y_a \p_u + \frac{v}{2}\p_a\right) \O(x)\,.
\end{equation}
The finite transformation generated by a null rotation is
\begin{align}\label{finiteN}
    e^{i\vec{\alpha}\cdot \vec{N}}\O(x)e^{-i\vec{\alpha} \cdot \vec{N}} = \O(\Lambda_{-\vec{\alpha}}x)\ , \qquad
    \Lambda_{-\vec{\alpha}}(u,v,\vec{y}) = \left(u+\vec{\alpha}\cdot \vec{y}+\frac{v}{4}\vec{\alpha}^{\,2},\,v,\,\vec{y}+\frac{v}{2}\vec{\alpha}\right).
\end{align}

\subsubsection{Relation to the detector frame}
In the Hofman--Maldacena conformal collider bounds \cite{Hofman:2008ar} and in particle physics \cite{Basham:1977iq,Basham:1978zq,Basham:1978bw,Basham:1979gh, Moult:2025nhu}, the ANEC operator is typically inserted at future null infinity. This models an energy calorimeter very far from the experiment. The ANEC through the origin can be mapped to the detector operator at null infinity by the  inversion
\begin{equation}
    v' = -\frac{1}{v}\,, \qquad\quad u' = u-\frac{\vec{y}^{\,2}}{v}\,, \qquad\quad \vec{y}^{\,\prime} = \frac{\vec{y}}{v}\,.
\end{equation}
The original metric in these coordinates is
\begin{equation}
    ds^2 = \frac{1}{(v')^2}\left(-du'\lsp dv'+d\vec{y}^{\,\prime\lsp 2}\right),
\end{equation}
and the null rotations are simply the transverse translations, $\zeta_a = \frac{1}{2}\p_{y^{\prime\lsp a}}$. In a conformal theory, a Weyl transformation eliminates the singular conformal factor, but in a non-conformal QFT the two setups are inequivalent. We will place the operator through the origin, for two reasons. First, the contact terms play an essential role and it is inconvenient to have contact terms at infinity. Second, our ANEC operator must measure energies at all scales, whereas a detector operator strictly at null infinity only detects massless particles. It could be interesting to study detectors at finite distance.

\subsection{Spectral function in impact parameter space}

Let's briefly recall the spectral decomposition of a scalar 2-point function, $\langle \O(x)\O(0)\rangle$. The spectral function $\rho_{\O\O}(s)$ is defined by inserting a momentum projector,
\begin{align}\label{oospectral}
    \theta(p^0)\rho_{\O\O}(-p^2) &= (2\pi)^3 \langle \O\, \delta^{(4)}(P-p)\, \O \rangle \\
    &= (2\pi)^3 \sum_X \delta^{(4)}(p-p_X)| \langle 0|\O|X\rangle|^2\,,
\end{align}
where $\O = \O(x=0)$. In the first line, $P$ is an operator, and the projector is the operator-valued delta function $\delta^{(4)}(P-p) = \int \frac{d^{\lsp 4} x}{(2\pi)^4} e^{ix \cdot(P-p)} $. To relate \eqref{oospectral} to the Wightman function, insert the integral representation of the projector,
\begin{align}
    (2\pi)^4 \langle \O\, \delta^{(4)}(P-p)\, \O \rangle &= \int d^{\lsp 4}x\, e^{-ip\cdot x}\langle \O(0)\, e^{iP\cdot x}\, \O(0)\rangle =\int d^{\lsp 4}x\, e^{-ip\cdot x}\braket{\O(x)\O(0)}\,.
\end{align}
Now we will repeat this with an ANEC insertion, $\langle \O \E \O\rangle$. Define the states
\begin{equation}\label{defOpm}
    |\O^\pm(\omega)\rangle = \int d^{\lsp 4} x \, \theta(\pm v) e^{-i\omega u+i\epsilon v}\O(x)|0\rangle\,, \qquad \omega > 0\,.
\end{equation}
The reason for including $\theta(\pm v)$ is that such states appear naturally in time-ordered correlators, as we will see below. The $i\epsilon$ in the exponent is a regulator that will be taken to zero and is included for the following reason. With this regulator, the state
\begin{equation}
    |\O^+(\omega)\rangle + |\O^-(\omega)\rangle = \int d^{\lsp 4}x\, e^{-i\omega u + i \epsilon v}\O(x)|0\rangle
\end{equation}
has spacelike momentum. Therefore it vanishes, so that
\begin{equation}\label{stateflip}
    |\O^+(\omega)\rangle = - |\O^-(\omega)\rangle\,.
\end{equation}
This can also be understood as the contour rotation $\int_0^\infty dv \to -\int_{-\infty}^0dv$.

The ANEC spectral function is defined by inserting a projector onto definite null rotations, with eigenvalue $n_a$:
\begin{equation}\label{defrhobEversion}
    \langle \O^-(\omega)| \E \delta^{(2)}(\vec{N} - \vec{n}) |\O^-(\omega')\rangle = 2\pi \delta(\omega-\omega') \rhoE_{\O\O}(\omega,\vec{n})\,.
\end{equation}
Equivalently,
\begin{equation}\label{defrhob1}
  \rhoE_{\O\O}(\omega, \vec{n}) = \langle \O^-(\omega)| T_{uu}(0) \delta^{(2)}(\vec{N}-\vec{n}) |\O^-(\omega)\rangle\,.
\end{equation}
Under a boost, $(u,v) \to (e^\gamma u, e^{-\gamma} v)$, the density $\rhoE_{\O\O}$ is invariant and its arguments transform as $(\omega,\vec{n}) \to (e^{-\gamma}\omega,e^{-\gamma}\vec{n})$. It is also rotationally invariant. It follows that we can write it as a function of the impact parameter $b = |\vec{b}| = |\vec{n}|/\omega$:
\begin{equation}\label{defrhob}
    \rhoE_{\O\O}(b) = \rhoE_{\O\O}(\omega, \omega \vec{b}) = \langle \O^-(\omega)| T_{uu}(0) \delta^{(2)}(\vec{N}-\omega \vec{b}) |\O^-(\omega)\rangle\,.
\end{equation}
The spectral function is manifestly non-negative by the ANEC,
\begin{equation}
    \rhoE_{\O\O}(b) \geq 0\,,
\end{equation}
pointwise in $\vec{b}$. This discussion easily generalizes to multiple operators $\O_I$, with spectral functions $\rhoE_{IJ}$, which give a positive semidefinite matrix in the indices $I,J$.

In a CFT, it is not necessary to do a spectral decomposition to extract the complete set of positivity conditions from the 3-point functions involving $\E$ \cite{Hofman:2008ar}. This is because the 3-point functions are determined by conformal invariance up to several constant coefficients. In a QFT, the ANEC 3-point correlator is positive at all scales, and what we have done is to organize this positivity by impact parameter.

The spectral function can also be written as a smeared Wightman correlator using \eqref{finiteN}. Using
\begin{equation}
    \delta^{(2)}(\vec{N}-\omega \vec{b}) = \int \frac{d^{\lsp 2}\vec{\alpha}}{(2\pi)^2} e^{i \vec{\alpha}\cdot(\vec{N}-\omega \vec{b})}\,,
\end{equation}
applied to \eqref{defrhobEversion}, with $e^{i\vec{\alpha} \cdot \vec{N}}$ split equally between the bra and ket, gives
\begin{align}
    \rhoE_{\O\O}(b) &= \int\frac{d^{\lsp 2}\vec{\alpha}}{(2\pi)^2}\,e^{-i\omega\vec{\alpha}\cdot \vec{b}} \langle\O^-(\omega)|e^{i\vec{\alpha}\cdot \vec{N}/2} T_{uu}(0)e^{i\vec{\alpha}\cdot \vec{N}/2}|\O^-(\omega)\rangle \notag\\
    &= \frac{4i}{\pi\omega}\int_{v_1,v_2<0} \frac{d^{\lsp 4}x_1\,d^{\lsp 4}x_2}{v_1-v_2+i\epsilon}\, e^{i\omega(u_1-u_2)-i\epsilon(v_1-v_2)}  e^{-i\frac{\omega(\vec{y}_1+\vec{y}_2-2\vec{b})^2}{(v_1-v_2+i\epsilon)}} \langle\O(x_1)T_{uu}(0)\O(x_2)\rangle\,.
\end{align}

\subsection{Relation to the time-ordered correlator}\label{ss:relationTO}
We use the notation
\begin{equation}
    E_{uu}(k) = T_{uu}(k)\big|_{k_u = 0}\,.
\end{equation}
This is the Fourier transform of $\E(v,\vec{y})$, and related to the ANEC through the origin by
\begin{equation}
\mathcal{E} 
= \int \frac{dk_v\, d^2\vec{k}}{\pi(2\pi)^2}E_{uu}(k)\, .
\end{equation}
We will study the time-ordered correlation function
\begin{equation}\label{GEtodef}
    \GE_{\O\O}(s_1,s_2;t) = \omega^{-2}\llangle {\cal T}[ \O(k_1) \O(k_2) E_{uu}(k_3)]\rrangle\,,
\end{equation}
where we have defined
\begin{equation}
    s_1 = -k_1^2\,, \qquad s_2 = -k_2^2\,, \qquad t = - (k_1+k_2)^2\,, \qquad k_{1u} = -k_{2u} = \omega >0\,.
\end{equation}
Since $k_{3u} = 0$, the momentum transfer is 
\begin{equation}
    t = - (k_3)^2 = - \vec{k}_3^{\,2}\,.
\end{equation}
The 3-point function in impact parameter space is denoted $\FE_{\O\O}$ and defined by the transverse Fourier transform,
\begin{equation}
    \FE_{\O\O}(s_1,s_2;b) = \int \frac{d^{\lsp 2} \vec{q}}{(2\pi)^2} e^{-i\vec{q} \cdot \vec{b}}\GE_{\O\O}(s_1,s_2;t = -\vec{q}^{\,2})\,.
\end{equation}
We will now derive the crucial relation between the time-ordered correlator and the spectral density:
\begin{equation}\label{FErho}
    \FE_{\O\O}(0,0;b) = -\rhoE_{\O\O}(b)\, , \qquad  \text{ for }\vec{b} \neq 0\,.
\end{equation}
Therefore, for null $k_1$ and $k_2$, $\FE_{\O\O}(0,0;b) \leq 0$ except at the origin, where there can be contact terms of either sign. In \eqref{FErho}, accounting for the $\epsilon$ in \eqref{defOpm} requires defining the null limit from the spacelike direction, $\FE_{\O\O}(0,0;b) = \lim_{s_1, s_2 \to 0^-} \FE_{\O\O}(s_1,s_2;b)$.

The surprisingly simple relation \eqref{FErho} between the time-ordered correlator and the spectral density is a special feature of the ANEC operator. It comes from two simplifications. The first is that $\E|0\rangle = 0$ eliminates many of the terms in the time-ordered correlator, leaving only the Wightman function $\langle \O \E \O \rangle$ plus contact terms \cite{Hartman:2023qdn,Hartman:2024xkw}. The second is that in impact parameter space, all contact terms in this correlation function --- including semi-local or partial contact terms with two operators coincident and the third operator separated --- sit at $\vec{b}=0$.

\subsubsection{Derivation of \eqref{FErho}}

Define
\begin{align}
    |\O^-(\omega,\vec{q})\rangle &= e^{i\vec{q}\cdot \vec{N} / \omega}|\O^-(\omega)\rangle \\
    &= \int d^{\lsp 4} x\, \theta(-v) e^{-i\omega u+i\epsilon v + i \vec{q}\cdot \vec{y} - i \frac{\vec{q}^{\,2} v}{4\omega}}\O(x)|0\rangle\,,
\end{align}
where we used \eqref{defOpm}.
The time-ordered correlator at the regulated spacelike momenta is
\begin{equation}\label{towight}
    \GE_{\O\O}(-\epsilon,-\epsilon,t=-\vec{q}^{\,2}) = \frac{1}{\omega^2} \langle \O^+(\omega)| T_{uu}(0) e^{i\vec{q}\cdot \vec{N}/\omega} |\O^-(\omega)\rangle + \text{contact}\,,
\end{equation}
where we used $\E |0\rangle = 0$ and $\omega > 0$ to eliminate the other orderings. We will show momentarily that contact terms are supported only at $\vec{b}=0$. Therefore, for $\vec{b} \neq 0$,
\begin{align}
    \FE_{\O\O}(0,0;b) &= \frac{1}{\omega^2} \int \frac{d^{\lsp 2}\vec{q}}{(2\pi)^2} e^{-i\vec{q}\cdot \vec{b}} \langle \O^+(\omega)| T_{uu}(0) e^{i\vec{q}\cdot \vec{N}/\omega}|\O^-(\omega)\rangle \\
    &=\langle \O^+(\omega)| T_{uu}(0) \delta^{(2)}(\vec{N}-\omega \vec{b}) | \O^-(\omega)\rangle\\
    &= -\rhoE_{\O\O}(b)\,.
\end{align}
In the last line we used \eqref{stateflip} and \eqref{defrhob}.

Now consider the contact terms. There are several different kinds of contact terms in the three-point function $\langle {\cal T}[\O(x_1) \E \O(x_2)]\rangle$. First, there are local terms where all three operators coincide. These are obviously at $\vec{b}=0$. Second, there are partial contact (semi-local) terms where $x_1=x_2$, separated from the ANEC. These vanish using $\E|0\rangle = 0$. The troublesome partial contact terms are the ones where $T_{uu}$ coincides with one of the $\O$'s, with the second $\O$ separated. These partial contact terms were the source of considerable technical difficulty in \cite{Hartman:2023qdn}. Fortunately, they are all localized at $\vec{b}=0$ in the impact parameter representation of the correlator. This is intuitively clear from the kinematics --- an operator $\O$ cannot coincide with $\E$ at nonzero impact parameter --- and can be established as follows. The problematic partial contact terms have the form
\begin{equation}
    \llangle {\cal T}[ \O(k_1) T_{\mu\nu}(k_3) \O(k_2) ]\rrangle_{\text{PC}} = \sum_{\mathcal X} P^{\mathcal X}_{\mu\nu}{}^{\alpha\beta\dots} (k_3) \llangle {\cal T}[ \mathcal X_{\alpha\beta\dots}(-k_2) \O(k_2)] \rrangle + (k_1 \leftrightarrow k_2)\,,
\end{equation}
where the sum is over local operators $\mathcal X$ and $P$ is a polynomial. In terms of the invariants, the most general polynomial that we can write is
\begin{equation}
    \GE_{\O\O}(s_1,s_2;t)_{\text{PC}} = \sum_{a,b \geq 0} t^a [s_1^b F_{ab}(s_2) + s_2^b F_{ab}(s_1) ]\,,\label{eq:GOOEPC1}
\end{equation}
where this is a finite sum with integer powers. This is a polynomial in $t$, which means in impact parameter space it only has support at $\vec{b}=0$.
Fourier transforming \eqref{eq:GOOEPC1} to impact parameter space gives
\begin{align}
      \FE_{\O\O}(s_1,s_2;b)_{\text{PC}}= \int \frac{d^{\lsp 2} \vec{q}}{(2\pi)^2} e^{-i\vec{q} \cdot \vec{b}}\GE_{\O\O}(s_1,s_2;t=-\vec{q}^{\,2})_{\text{PC}}
       =\sum_{a,b\geq 0}[s_1^bF_{ab}(s_2)+s_2^bF_{ab}(s_1)]\partial^{2a}\delta^{(2)}(\vec{b}) \, .
\end{align}

\subsubsection{Momentum space spectral decomposition} Up until this point, we have considered spectral representations in impact parameter space, as this is the natural space in which the ANEC provides pointwise positivity. Nevertheless, we can derive an interesting relation between the momentum-space spectral kernel and a double discontinuity of a correlator. The momentum space spectral function is defined as 
\begin{equation}\label{momentumspectralmt}
    \rhoE_{\O\O}(s_1, s_2; t) = (2\pi)^5 \langle \O| \delta^{(2)}(\vec{P} - \vec{q}_1) \delta(P^2+s_1) \E \delta(P^2+s_2) \delta^{(2)}(\vec{P} - \vec{q}_2) \delta(P_u+\omega) |\O\rangle\,,
\end{equation}
where multiple insertions of the projectors are necessary because $[P^2,\mathcal{E}]\neq0$, and where $t= -(\vec{q}_1-\vec{q}_2)^2$. This spectral kernel is proportional to the double discontinuity of the time-ordered correlator,
\begin{equation}
    \text{dDisc}\,  G_{\O\O}^{\E} (s_1,s_2;t) = 16\pi^2\rho_{\O\O}^\E(s_1,s_2;t)\, ,
\end{equation}
where $G_{\O\O}^\E$ is defined in \eqref{GEtodef}. This relation is derived in appendix \ref{app:momentum}.

\section{The non-perturbative positive metric}\label{s:exacta}

Define the state
\begin{equation}\label{defpsia}
    |\psi_a(\omega)\rangle = \omega \int d^{\lsp 4} x \, \theta(-v) e^{-i\omega u + i \epsilon v} y_a T(x)|0\rangle\,,
\end{equation}
where $T = T^\mu{\!}_\mu$. In \cite{Hartman:2023qdn} this state was used to show that the ANEC implies the $a$-theorem for RG flows in four dimensions. The key identity is the sum rule
\begin{equation}\label{oursumrule}
    \langle \psi_a(\omega)| \E | \psi_a(\omega')\rangle = 64\pi\omega^4\delta(\omega - \omega') \Delta a\, .
\end{equation}
This proves that the total change $\Delta a = a_{\text{UV}} - a_{\text{IR}}$ along an RG flow is positive, but says nothing about $a$ along the flow. We will now upgrade this result to a monotonic $a$-function along the flow using the results of section \ref{s:spectral}. The state defined in \eqref{defpsia} is related to the one in \eqref{defOpm}, for the operator $\O=T$, by the action of the null rotation generator,
\begin{equation}
    |\psi_a(\omega)\rangle = N_a |T^-(\omega)\rangle\,.
\end{equation}
Therefore, equation \eqref{oursumrule} can be rewritten as
\begin{equation}
    \delta(\omega - \omega')\Delta a = \frac{1}{64\pi\omega^4} \langle T^-(\omega)| N_a \E N_a |T^-(\omega')\rangle\,.
\end{equation}
Now to perform a spectral decomposition we insert the identity in the form
\begin{equation}
    \mathbf{1} = \omega^2 \int d^{\lsp 2} \vec{b} \, \delta^{(2)}(\vec{N} - \omega \vec{b})\,,
\end{equation}
and express the result in terms of the spectral function \eqref{defrhob} (see also \eqref{defrhobEversion}).  This gives the sum rule in the form
\begin{equation}\label{newsumrule}
    \Delta a = \frac{1}{32} \int d^{\lsp 2}\vec{b} \, \vec{b}^{\,2} \rhoE_{TT}(b)\,,
\end{equation}
where $\rhoE_{TT}(b)$ is given by \eqref{defrhobEversion} with $\O \to T$. Since $\rhoE_{TT}(b) \geq 0$, this expresses $\Delta a$ as a sum of positive contributions at each scale. Now we can define a corresponding monotonic $a$-function as a function of the RG scale $\mu$,
\begin{equation}\label{npafunction}
    \tilde{a}(\mu) = a_{\text{IR}} + \frac{1}{32}\int_{|\vec{b}| > \mu^{-1}} d^{\lsp 2} \vec{b} \, \vec{b}^{\,2} \rhoE_{TT}(b)\,.
\end{equation}
We fixed the integration constant to match the infrared anomaly in the $\mu\rightarrow 0$ limit, such that this flows monotonically from $a_{\text{IR}}$ in the infrared to $a_{\text{UV}}$ in the ultraviolet. The scale derivative is
\begin{equation}\label{npaflow}
    \mu \frac{d}{d\mu} \tilde{a} = \frac{\pi}{16\mu^4}\rhoE_{TT}(\mu^{-1})\geq0\,.
\end{equation}

In the form \eqref{newsumrule}, the four-dimensional result for $\Delta a$ is strikingly similar to Zamolodchikov's result in two dimensions, $\Delta c = 3\pi \int_{|x|\neq 0} d^{\lsp 2} x\, x^2 \langle T(x)T(0)\rangle$. By formulating the problem in terms of the impact parameter we have reduced the kinematics to the 2-dimensional transverse plane, where it mimics a two-dimensional CFT.

These results are exact and non-perturbative. Under some additional assumptions, we can use them to define a metric on the space of couplings of the QFT. Suppose we can expand the trace of the stress tensor in terms of a set of relevant (and marginally relevant) operators $\O_I$ as
\begin{equation}\label{texpansion}
    T = \beta^I \O_I\,,
\end{equation}
where $\beta^I$ are the beta functions for the couplings conjugate to $\O_I$. This expression does not require the operators to be marginal, or the beta functions to be small --- it is defined at a point along the flow, and the $\O_I$ do not need to be operators in the UV CFT. The beta functions, which in general are dimensionful, are defined by \eqref{texpansion}.\footnote{In particular, in a perturbative theory with global symmetries, they are the full, flavor-corrected beta functions often denoted $B^I$ in the literature \cite{Jack:1990eb,Fortin:2012hn,Baume:2014rla,Herren:2021yur}.} In the language of the effective action, \eqref{texpansion} says that the theory can be made Weyl invariant through the introduction of a dilaton, as in \cite{Komargodski:2011vj}, so this is a reasonable assumption non-perturbatively. There are no contact terms in \eqref{npaflow}, so we can freely replace $T \to \beta^I\O_I$ in the spectral density. We obtain
\begin{equation}\label{npaflowBeta}
    \mu \frac{d}{d\mu} \tilde{a} = \frac{\pi}{16\mu^4}\rhoE_{IJ}(\mu^{-1})\beta^I \beta^J\,,
\end{equation}
where $\rhoE_{IJ}(b)$ is the spectral density for $\langle \O_I \E \O_J\rangle$. This spectral density is positive semidefinite,
\begin{equation}
    \rhoE_{IJ}(b) \succeq 0\,.
\end{equation}
Thus we interpret $\rhoE_{IJ}$ as the metric on the space of theories that governs the RG flow.

\subsection{Improvement terms}
The sum rule derived in \cite{Hartman:2023qdn} is insensitive to stress tensor improvements. The same is true of the spectral density, $\rhoE_{TT}(b)$, for any $\vec{b} \neq 0$. (See also \cite{Nakamura:2025uyd} for a 2d discussion.) Consider an improvement term added to the stress tensor,
\begin{equation}
    T'_{\mu\nu} = T_{\mu\nu}+\xi (\p_\mu \p_\nu - \eta_{\mu\nu}\p^2)L\,, \qquad T' = T - 3 \xi \p^2 L\,.
\end{equation}
The ANEC is unchanged, as it is shifted by the total derivative $\xi \int du\, \p_u^2 L = 0$. The state $|T^-(\omega)\rangle$ shifts by
\begin{equation}
    |T^{\prime\lsp-}(\omega)\rangle = |T^-(\omega)\rangle + 6 i \omega \xi \int du\, d^{\lsp 2}\vec{y} \, e^{-i\omega u}L(u,0,\vec{y}) |0\rangle\,,
\end{equation}
where we integrated by parts to move $\p_v L$ onto the null plane $v=0$. For $\vec{b}\neq 0$, the operator in the second term commutes with the ANEC by causality, and therefore using the fact that the ANEC annihilates the vacuum, we find
\begin{equation}
    \E |T^{\prime\lsp-}(\omega)\rangle = \E | T^-(\omega)\rangle\,.
\end{equation}
This implies that away from $\vec{b}=0$ the spectral density $\rhoE_{TT}(b)$ is unchanged by improvements.

\section{Setup of the local renormalization group}\label{s:localrgsetup}

In the rest of the paper, we will consider perturbative RG flows, where the relevant operators $\O_I$ are classically marginal and the beta functions are small. In that context, we will reproduce the flow equation \eqref{npaflowBeta} from the local RG formalism of Jack and Osborn \cite{Osborn:1989td,Jack:1990eb,Osborn:1991gm}.

The setting of the local RG is a renormalizable QFT in $d$ dimensions with classically marginal operators $\O_I$. We assume the theory can be coupled to a general background metric $g_{\mu\nu}(x)$ and spacetime-dependent couplings $\lambda^I(x)$. To keep the formalism simple, we also assume there are no classically relevant operators, but we do not expect this to affect the conclusions. The local RG is a perturbative framework, so the discussion is limited to theories in which the beta functions and their derivatives are small, and remain small along the entire flow. This does not necessarily mean the theory itself is weakly coupled; it also applies to short flows between interacting fixed points.

A theory with nonzero beta functions is invariant under a combined local Weyl rescaling and coupling constant transformation,
\begin{equation}
    \delta g^{\mu\nu} = 2\sigma g^{\mu\nu}\,, \qquad \delta \lambda^I= -\sigma \beta^I\,.\label{eq:4p1}
\end{equation}
Let
\begin{equation}\label{eq:partitionFunction}
    Z(\lambda,g)=\int D\phi\, e^{-S[\phi;\lsp g,\lambda]}\,,
\end{equation}
and define the generators
\begin{equation}
    \Delta_\sigma^g = 2 \int d^{\lsp d} x \, \sigma g^{\mu\nu} \frac{\delta}{\delta g^{\mu\nu}}\,, \qquad \quad\Delta_\sigma^\lambda = \int d^{\lsp d}x \, \sigma \beta^I \frac{\delta}{\delta \lambda^I}\,,
\end{equation}
which act as
\begin{equation}
    \Delta_\sigma^g \log Z = - \int d^{\lsp d} x \sqrt{g} \, \sigma g^{\mu\nu}\langle T_{\mu\nu} \rangle_{g,\lambda}\,, \qquad \Delta_\sigma^\lambda \log Z = - \int d^{\lsp d} x \sqrt{g} \, \sigma \beta^I \langle \O_I \rangle_{g,\lambda}\,.
\end{equation}
The combined variation is
\begin{equation}\label{combovar}
    \Delta_\sigma \log Z = - \int d^{\lsp d} x \sqrt{g}\, \sigma \langle \opS \rangle_{g,\lambda}\,, \qquad \Delta_\sigma = \Delta_\sigma^g - \Delta_\sigma^\lambda\,,
\end{equation}
where we have defined the operator
\begin{equation}
    \opS = T - \opTheta\,, \qquad T = g^{\mu\nu}T_{\mu\nu}\,, \qquad \opTheta = \beta^I \O_I\,.
\end{equation}
The statement that the theory is invariant under \eqref{eq:4p1} means that the variation \eqref{combovar} is purely local, since the violation of invariance can only arise from counterterms, which are local. That is, $\opS =0$ as an operator equation, and
\begin{equation}\label{anomalyEquation}
    \langle \opS \rangle_{g,\lambda} = {\cal A}_d\,,
\end{equation}
where ${\cal A}_d$ is a local scalar built from $g_{\mu\nu}(x)$ and $\lambda^I(x)$. Equation \eqref{anomalyEquation} is the generalized Weyl anomaly, and at a conformal fixed point it reduces to the usual formula. Since $\opS$ has dimension $d$, the right-hand side vanishes in odd dimensions in a theory with no classical scale. 

Our conventions for correlation functions, defined by varying $\log Z$, are described in appendix \ref{app:corr}. It is evident from \eqref{anomalyEquation} that all correlation functions involving $\opS$ are pure contact terms, since they can be obtained by varying the local quantity ${\cal A}_d$. Note that we define correlation functions of $T$ and $\opTheta$ with the metric contraction and the beta functions inside the variations, which means that $g^{\mu\nu}T_{\mu\nu}$ and $T$ are identical as operators but their correlation functions differ by contact terms, and similarly for $\opTheta$ and $\beta^I\O_I$.

A change in the RG scale $\mu$ can be absorbed into a change of units, which corresponds to the trivial equation
\begin{equation}
    \left( \mu \frac{\p}{\p \mu} + 2 \int d^{\lsp d} x\, g^{\mu\nu} \frac{\delta}{\delta g^{\mu\nu}} \right) \log Z = 0\,.
\end{equation}
Using this relation together with \eqref{combovar} at $\sigma = 1$ we obtain
\begin{equation}
    \left(\mu \frac{\partial}{\partial \mu} + \int d^{\lsp d}x\, \beta^I \frac{\delta}{\delta\lambda^I}\right)\log{Z} = \int d^{\lsp d}x\sqrt{g}\,\mathcal{A}_d\,.\label{eq:GeneralizedCS}
\end{equation}
This is the local Callan--Symanzik equation. In a flat background and with spacetime-independent couplings, it becomes the ordinary Callan--Symanzik equation
\begin{equation}\label{DlogZ}
    \D\log{Z} = 0\,,
\end{equation}
where the total RG derivative is
\begin{equation}
    \mu \frac{d}{d\mu} = \mu \frac{\partial}{\partial\mu}+ \beta^I\frac{\partial}{\partial\lambda^I}\,, \qquad \beta^I = \mu \frac{d}{d\mu} \lambda^I\,.\label{eq:defD}
\end{equation}
The RG equations for correlation functions are obtained by varying \eqref{eq:GeneralizedCS}. For example,
\begin{equation}\label{eq:DTuvTrs2d}
    \D\braket{T_{\mu\nu}(x_1)T_{\rho\sigma}(x_2)} = \frac{4}{\sqrt{ g(x_1)}\sqrt{g(x_2)}}\frac{\delta}{\delta g^{\mu\nu}(x_1)}\frac{\delta}{\delta g^{\rho\sigma}(x_2)}\int d^{\lsp d}x \sqrt{g}\,\mathcal{A}_d\,.
\end{equation}
In expressions such as this one with a correlator in the QFT without sources on the left-hand side, it is implicit that the right-hand side is evaluated in a flat metric with constant $\lambda^I$. The derivative $\D$ commutes with $g^{\mu\nu}$, $\frac{\delta}{\delta g^{\mu\nu}}$, and $\beta^I \frac{\delta}{\delta \lambda^I}$, while
\begin{equation}\label{Dlambdacomm}
    \left[\D, \frac{\delta}{\delta\lambda^I(x)}\right] = - \gamma_I{}^J \frac{\delta}{\delta \lambda^J(x)}\,,
\end{equation}
with $\gamma_I{}^J = \p_I \beta^J$ the anomalous dimension. In momentum space, acting with $\D$ is related to inserting $\opS$ at zero momentum. The example \eqref{eq:DTuvTrs2d} becomes
\begin{equation}
    \D \llangle T_{\mu\nu}(k) T_{\rho\sigma}(-k) \rrangle = \llangle T_{\mu\nu}(k) T_{\rho\sigma}(-k) \opS(0) \rrangle\,,
\end{equation}
where $\llangle \dots \rrangle$ is the momentum-space correlation function with the delta function removed (see \eqref{eq:DoubleBrackets}).

\section{Local RG in two dimensions}\label{s:localrg2d}

In this section, we review the application of the local RG formalism to two-dimensional QFTs. All of the results in this section are present in the literature \cite{Osborn:1991gm} in some form, but the details of how we organize the flow equations and the choice of renormalization scheme are a bit different. We will use the local RG to rederive the Zamolodchikov $c$-function, in a way that generalizes naturally to four dimensions by inserting an additional ANEC operator.

\subsection{Generalized anomaly}
In two dimensions, the generalized Weyl anomaly takes the form \cite{Osborn:1991gm}
\begin{equation}
    \mathcal{A}_2 = \tfrac12C R-\tfrac12\chi_{IJ}\partial_\mu \lambda^I\partial^\mu \lambda^J + \nabla_\mu (w_I\partial^\mu \lambda^I)\,,\label{eq:A2}
\end{equation}
where $C$, $\chi_{IJ}$, and $w_I$ are functions of the couplings $\lambda^I(x)$, and $R$ is the Ricci scalar. This is the most general local quantity of dimension two. At a conformal fixed point, with constant $\lambda^I$, the central charge of the CFT is $c = 12 \pi C$ and this reduces to the standard Weyl anomaly.

The trace is scheme-dependent, and the free functions $C,\chi_{IJ}$, and $w_I$ can be shifted by adding counterterms. A general local counterterm in two dimensions is
\begin{equation}\label{t:logZct}
    \log Z_{\rm ct} = \int d^{\lsp 2} x \, \sqrt{g} \left(\tfrac12 b(\lambda)R-\tfrac12 c_{IJ}(\lambda)\partial_\mu\lambda^I\partial^\mu\lambda^J\right).
\end{equation}
Adding this term shifts the anomaly functions appearing in \eqref{eq:A2} by
\begin{equation}\label{t:countershifts}
    \begin{aligned}
        C &\to C+\beta^I\p_I b\,,\\
        w_I &\to w_I-\p_I b+c_{IJ}\beta^J\,,\\
        \chi_{IJ} &\to \chi_{IJ}+\mathsterling_{\!\beta}c_{IJ}\,,
    \end{aligned}
\end{equation}
where $\mathsterling_{\!\beta}c_{IJ} = \beta^K\p_K c_{IJ} +(\p_I\beta^K)c_{KJ}+(\p_J\beta^K)c_{IK}$ is the Lie derivative in flavor space along the vector field $\beta^I$. The functions $C,\chi_{IJ}$, and $w_I$ have no explicit dependence on the RG scale, $\mu$, but they depend on it through the running couplings $\lambda^I(\mu)$. As we will see, this condition is maintained under the shifts \eqref{t:countershifts} provided that the renormalization conditions are specified appropriately.

\subsection{Weyl consistency conditions}
The Wess--Zumino \cite{Wess:1971yu} consistency condition for the Weyl anomaly is the requirement that two successive Weyl transformations must commute,
\begin{equation}\label{sscom}
    \langle [\opS(x), \opS(y)] \rangle_{g,\lambda} = 0\,.
\end{equation}
This follows immediately from the definition of $\opS$-correlators as variations of $\log Z$, but it is not automatic from \eqref{eq:A2}, so it imposes relations among the functions $C$, $\chi_{IJ}$, and $w_I$. One way to derive these relations is to evaluate \eqref{sscom} directly. In integrated form,
\begin{align}\label{eq:wcc2d}
    [\Delta_{\sigma_1}, \Delta_{\sigma_2}]\log Z &= -\Delta_{\sigma_1}\int d^{\lsp 2} x \sqrt{g}\,\sigma_2 {\cal A}_2 +\Delta_{\sigma_2}\int d^{\lsp 2} x \sqrt{g}\,\sigma_1 {\cal A}_2  \\
    &=\int d^{\lsp 2}x\sqrt{g}\, \left(\sigma_2\partial_\mu\sigma_1-\sigma_1\partial_\mu\sigma_2\right)\left(\partial_I C+\beta^J\partial_J w_I+w_J\partial_I\beta^J -\chi_{IJ}\beta^J\right)\partial^\mu\lambda^I\,.\notag
\end{align}
Setting this to zero gives the consistency condition \cite{Osborn:1991gm}
\begin{equation}\label{flowTildeC}
    \partial_I\widetilde C =\left(\chi_{IJ}+\partial_I w_J-\partial_J w_I\right)\beta^J\,, \qquad \widetilde C= C+\beta^Iw_I\,.
\end{equation}
Contracting this equation with $\beta^I$ gives the flow equation
\begin{equation}\label{t:Cflow}
    \D\widetilde{C} = \chi_{IJ} \beta^I \beta^J\,.
\end{equation}

It is instructive to rephrase this derivation in terms of correlation functions of the flat-space, constant-coupling QFT. Varying \eqref{eq:A2} once with respect to the metric defines the correlator
\begin{equation}
    \braket{T_{\mu\nu}(x_1)\opS(x_2)} = -\frac{2}{\sqrt{g(x_1)}\sqrt{g(x_2)}}\frac{\delta}{\delta g^{\mu\nu}(x_1)}\left(\sqrt{g(x_2)}\mathcal{A}_2(x_2)\right)\, ,\label{eq:TuvS2ddef} 
\end{equation}
and similarly with respect to the coupling. Explicitly
\begin{align}\label{t:TddS}
    \langle T_{\mu\nu}(x) \opS(0)\rangle &= C (\p_\mu \p_\nu - \delta_{\mu\nu}\p^2 )\delta^{(2)}(x)\,,\\
    \langle \O_I(x)\opS(0)\rangle &= -w_I \p^2 \delta^{(2)}(x)\,.\label{t:OiS}
\end{align}
Using these equations to evaluate the combination  $TT - \opTheta\opTheta = (T+\opTheta)\opS$ gives
\begin{equation}\label{TTminusThTh}
    \langle T(x) T(0) \rangle - \langle \opTheta(x) \opTheta(0)\rangle = -\widetilde{C} \p^2 \delta^{(2)}(x)\,,
\end{equation}
where $\widetilde{C}$ was defined in \eqref{flowTildeC}. We want to act with $\D$ on this equation. First, we calculate the $\opS$-correlators by two variations of \eqref{eq:A2},
\begin{align}\label{t:TTS}
    \langle T(x_1) T(x_2) \opS(0) \rangle &= 0\,,\\
    \langle \opTheta(x_1) \opTheta(x_2) \opS(0) \rangle &= \begin{aligned}[t]
        &-\chi_{IJ}\beta^I\beta^J\, \p_{1\mu}\delta^{(2)}(x_1)\,\p_2^\mu\delta^{(2)}(x_2)\\
        &\qquad+\beta^I\p_I(\beta^Jw_J)\,(\p_1+\p_2)^2 \left[\delta^{(2)}(x_1)\delta^{(2)}(x_2)\right].\notag
    \end{aligned}
\end{align}
Now integrating over the position of the $\opS$ insertion gives the RG equations,
\begin{align}\label{t:DTT}
    \D \langle T(x) T(0) \rangle &= 0\,,\\
    \D\langle \opTheta(x) \opTheta(0) \rangle &= \chi_{IJ}\beta^I\beta^J\,\p^2\delta^{(2)}(x)\,.\notag
\end{align}
Acting with $\D$ on \eqref{TTminusThTh} gives the flow equation \eqref{t:Cflow} for $\widetilde{C}$.

The uncontracted flow equation \eqref{flowTildeC} can be derived similarly. The other RG equations for one- and two-point functions obtained by varying the trace relation are
\begin{subequations}
    \begin{align}
        \D \langle T_{\mu\nu}\rangle &= 0\,,\\
        \D \langle \O_I \rangle &= -\p_I\beta^J\,\langle \O_J\rangle\,,\\
        \D \langle T_{\mu\nu}(x) T_{\rho\sigma}(0)\rangle &= 0\,,\\
        \D \langle T_{\mu\nu}(x) \O_I(0)\rangle &= -\p_I\beta^J\,\langle T_{\mu\nu}(x)\O_J(0)\rangle +\p_I C\,(\delta_{\mu\nu}\p^2-\p_\mu\p_\nu)\delta^{(2)}(x)\,,\\
        \D \langle \O_I(x) \O_J(0)\rangle &= -\p_I\beta^K\,\langle \O_K(x)\O_J(0)\rangle -\p_J\beta^K\,\langle \O_I(x)\O_K(0)\rangle\label{t:DOiOj}\nonumber\\[-5pt]
        &\qquad\qquad\qquad+\p_I\p_J\beta^K\,\langle \O_K\rangle\,\delta^{(2)}(x) +\chi_{IJ}\,\p^2\delta^{(2)}(x)\,.
    \end{align}
\end{subequations}
To derive these equations, it is important to keep track of the contact terms that arise from \eqref{Dlambdacomm}. Acting with $\D$ on \eqref{t:OiS} and using these relations reproduces \eqref{flowTildeC}.

\subsection{Integral relations}\label{ss:2dintegralrelations}
The correlation functions studied in the previous subsection can be used to write integral formulae for the anomaly functions, $C$, $w_I$, and $\chi_{IJ}$. As a first example, consider \eqref{TTminusThTh}. Multiplying this formula by $x^2$ and integrating over $x$, we find
\begin{align}\label{t:Cintegral}
    \widetilde{C} &= -\frac{1}{4}\int d^{\lsp 2} x \, x^2 \left( \langle T(x)T(0)\rangle - \langle \opTheta(x)\opTheta(0)\rangle\right) \\
    &= -\frac{1}{4}\int d^{\lsp 2} x \, x^2 \left( \langle T(x)T(0)\rangle - \beta^I \beta^J \langle \O_I(x) \O_J(0)\rangle \right),\notag
\end{align}
where in the second line we used $\int d^{\lsp 2} x\, x^2 \delta^{(2)}(x) =0 $ to eliminate a contact term. Similarly, we can invert \eqref{t:OiS} to solve for $w_I$,
\begin{equation}\label{eq:2dwformula}
    w_I = -\frac{1}{4}\int d^{\lsp 2} x\, x^2 \left( \langle \O_I(x)T(0)\rangle - \beta^J\langle \O_I(x) \O_J(0)\rangle \right),
\end{equation}
and integrating \eqref{t:DOiOj} gives the relation
\begin{align}
    \chi_{IJ} &= \frac{1}{4}\int d^{\lsp 2}x\, x^2 \left(\D\braket{\O_I(x)\O_J(0)}+\partial_I\beta^K\braket{\O_K(x)\O_J(0)}+\partial_J\beta^K\braket{\O_I(x)\O_K(0)}\right) \notag\\
    &= \frac{1}{4}\int d^{\lsp 2}x\, x^2 \mathsterling_{\!\beta} \braket{\O_I(x)\O_J(0)}\,.\label{eq:chiIJasLie}
\end{align}
Contracting with the beta functions,
\begin{equation}
    \chi_{IJ}\beta^I\beta^J = \frac{1}{4}\int d^{\lsp 2}x\, x^2 \D\big[\beta^I\beta^J \braket{\O_I(x)\O_J(0)}\big]\,.
\end{equation}
The correlators appearing in these integrals have explicit dependence on the RG scale $\mu$, while the functions $\widetilde{C}$, $w_I$, and $\chi_{IJ}$ do not; they are functions only of the running couplings. To understand why this is consistent, consider \eqref{t:Cintegral}.
By dimensional analysis, the correlators can be written in the form
\begin{equation}
    \langle T(x) T(0)\rangle = \frac{1}{x^4} F(\mu|x|, \lambda)\,, \qquad \langle \O_I(x)\O_J(0)\rangle = \frac{1}{x^4} G_{IJ}(\mu|x|, \lambda)\,.
\end{equation}
The explicit $\mu$ dependence of the integral can be removed by rescaling,
\begin{equation}
    \int d^{\lsp 2} x \, x^2 \langle T(x)T(0)\rangle = \int \frac{d^{\lsp 2}x}{x^2} F(\mu|x|, \lambda) = \int \frac{d^{\lsp 2}z}{z^2} F(|z|,\lambda)\,,
\end{equation}
where we introduced the dimensionless coordinate $z = \mu x$, and similarly for the other integral. This also justifies the derivation of \eqref{eq:chiIJasLie} from the line above; $\textit{a priori}$ the line above also includes $\mu \frac{\p}{\p \mu}$ acting on the correlator, but that contribution vanishes by scale invariance of the full integral.

\subsection{The Jack--Osborn \texorpdfstring{$c$}{c}-function}\label{ss:JOcsumrule}
The formula \eqref{t:Cintegral} defines the Jack--Osborn $c$-function, $\widetilde{C}(\lambda)$. The first term is a constant by the RG equation $\D\langle TT\rangle =0$ derived in  \eqref{t:DTT}, so
\begin{align}\label{JOcBeforeScheme}
    \widetilde{C} &= C_0 + \frac{1}{4}\beta^I \beta^J \int d^{\lsp 2} x \,  x^2 \langle \O_I(x)\O_J(0)\rangle \\
    &= C_0 + \frac{1}{4} \beta^I \beta^J \int \frac{d^{\lsp 2} z}{z^2} G_{IJ}(|z|, \lambda)\,,
\end{align}
where we again used the coordinate $z = \mu x$. We would like to evaluate this relation at the UV and IR fixed points. This is subtle: we cannot simply set $\beta^I = 0$ in this expression, because there are contributions to the integral that diverge as we approach a fixed point. Let us separate the last integral into two regions delimited by an arbitrary positive number, $s$. We have
\begin{equation}\label{t:Csplit}
    \widetilde{C} = C_0 + \frac{1}{4}\beta^I \beta^J \int_{|z|<s}\frac{d^{\lsp 2}z}{z^2} G_{IJ}(|z|,\lambda) + \frac{1}{4} \int_{|z|>s} \frac{d^{\lsp 2} z}{z^2} F(|z|,\lambda)\,,
\end{equation}
where in the region $|z|>s$ we have used $T = \beta^I \O_I$, which is allowed because this part of the integral has no coincident points. The operators $\O_I$ are marginally relevant at the UV fixed point, which implies that the first integral in \eqref{t:Csplit} converges near the origin (not at one loop, but after re-summing the anomalous dimensions). Consequently, that term can be dropped when a fixed point $\lambda_*$ is approached, since the vanishing beta functions multiply a finite number. Therefore, at a fixed point we have
\begin{equation}\label{t:CfixedInt}
    C(\lambda_*) = C_0 + \frac{1}{4}\lim_{\mu \to \mu_*} \int_{|x|>s/\mu} d^{\lsp 2}x \, x^2 \langle T(x)T(0)\rangle\,.
\end{equation}
This is now written in a form where we can safely evaluate it at either the UV or IR fixed point. At the IR fixed point, $\mu_* = 0$, the integral drops out and we learn that the constant term is in fact the IR value,
\begin{equation}
    C_0 = C_{\text{IR}} = \widetilde{C}(\lambda_{\text{IR}}) = \frac{c_{\text{IR}}}{12 \pi}\,,
\end{equation}
where $c_{\text{IR}}$ is the central charge of the infrared CFT. At the UV fixed point, the limit of integration in \eqref{t:CfixedInt} goes to zero, and we have
\begin{equation}
    C_{\text{UV}} = \widetilde{C}(\lambda_{\text{UV}}) = C_{\text{IR}} + \lim_{\epsilon \to 0} \frac{1}{4} \int_{|x|>\epsilon} d^{\lsp 2}x \, x^2 \langle T(x) T(0)\rangle\,.
\end{equation}
The difference is manifestly positive,
\begin{equation}
    C_{\text{UV}} - C_{\text{IR}} = \lim_{\epsilon \to 0} \frac{1}{4} \int_{|x|>\epsilon} d^{\lsp 2}x \, x^2 \langle T(x) T(0)\rangle\,,
\end{equation}
and this sum rule agrees with the one obtained by Zamolodchikov \cite{Zamolodchikov:1986gt,Cappelli:1990yc}.

\subsection{The Zamolodchikov \texorpdfstring{$c$}{c}-function\label{s:czamo}}
The Zamolodchikov $c$-function and its monotonic flow equation are obtained from the Jack--Osborn $c$-function by a specific choice of scheme.\footnote{A different method to define the Zamolodchikov-type scheme based on differential regularization was used in \cite{Osborn:1991gm}. It can be shown by direct integration that the scheme there satisfies \eqref{t:Zscheme}.} Recall the scheme dependence of the anomaly coefficients in equations \eqref{t:countershifts}. Let us define a scheme by imposing the renormalization condition
\begin{equation}\label{t:Zscheme}
    \int_{|x| < \mu^{-1}} d^{\lsp 2} x \, x^2 \langle \O_I(x) \O_J(0) \rangle = 0\,.
\end{equation}
The counterterm $c_{IJ}$ shifts the 2-point function by
\begin{equation}
    \langle \O_I(x)\O_J(0) \rangle \to \langle \O_I(x)\O_J(0)\rangle + c_{IJ}\,\p^2\delta^{(2)}(x)\,,
\end{equation}
hence we can always tune the counterterm to impose \eqref{t:Zscheme}. Denote the $c$-function in this scheme by $C_Z$. This function is related to the original, arbitrary reference scheme with $c$-function $\widetilde{C}$ by
\begin{equation}\label{czdef}
    C_Z(\lambda) = \widetilde{C}(\lambda) + c_{IJ}(\lambda) \beta^I(\lambda)\beta^J(\lambda)\,.
\end{equation}
In this scheme, equation \eqref{t:Csplit} with $s=1$ becomes
\begin{equation}\label{czint}
    C_Z = C_{\text{IR}} + \frac{1}{4} \int_{|x|>\mu^{-1}} d^{\lsp 2}x \, x^2 \langle T(x) T(0)\rangle\,.
\end{equation}
This is exactly the Zamolodchikov $c$-function, and it satisfies the monotonic flow equation
\begin{equation}\label{CZflow}
    \mu \frac{d}{d\mu} C_Z = \frac{\pi}{2} \mu^{-4} \langle T(\mu^{-1}) T(0) \rangle \geq 0\,.
\end{equation}
The $\mu$ derivative acts only on the endpoint of integration in \eqref{czint} because $\D\langle TT\rangle =0$.

It is not obvious, but $C_Z$ depends on $\mu$ only through the running coupling, i.e.\ $\mu \frac{\p}{\p \mu} C_Z = 0$. The key point is that \eqref{t:Zscheme} is a scale-invariant integral. In more detail, let us denote the correlator in the reference scheme by $\langle \O_I\O_J\rangle_{\rm ref}$. Then the condition \eqref{t:Zscheme} is imposed by choosing the finite counterterm
\begin{equation}
    c_{IJ} = - \frac{1}{4} \int_{|x| <\mu^{-1}} d^{\lsp 2}x\, x^2 \langle \O_I(x)\O_J(0)\rangle_{\rm ref} = - \frac{1}{4} \int_{|x| < \mu^{-1}} \frac{d^{\lsp 2} x}{x^2}\, G_{IJ}^{\rm ref}(|x|\mu, \lambda)\,.
\end{equation}
Rescaling the integration variable $x = z \mu^{-1}$ eliminates all of the explicit $\mu$ dependence.

By re-expressing  \eqref{CZflow} in terms of the marginal operators we can see the flow equation \eqref{t:Cflow} and corresponding positive metric. At separated points we can replace $T = \beta^I\O_I$, so the flow equation is
\begin{align}
    \mu \frac{d}{d\mu}C_Z &= \frac{\pi}{2}\mu^{-4}\langle \O_I(\mu^{-1})\O_J(0)\rangle\,\beta^I\beta^J \notag\\
    &= \frac{\pi}{2}G_{IJ}(1,\lambda)\,\beta^I\beta^J\,,
\end{align}
with $G_{IJ}$ the positive-definite Zamolodchikov metric.

\subsection{Momentum-space version}

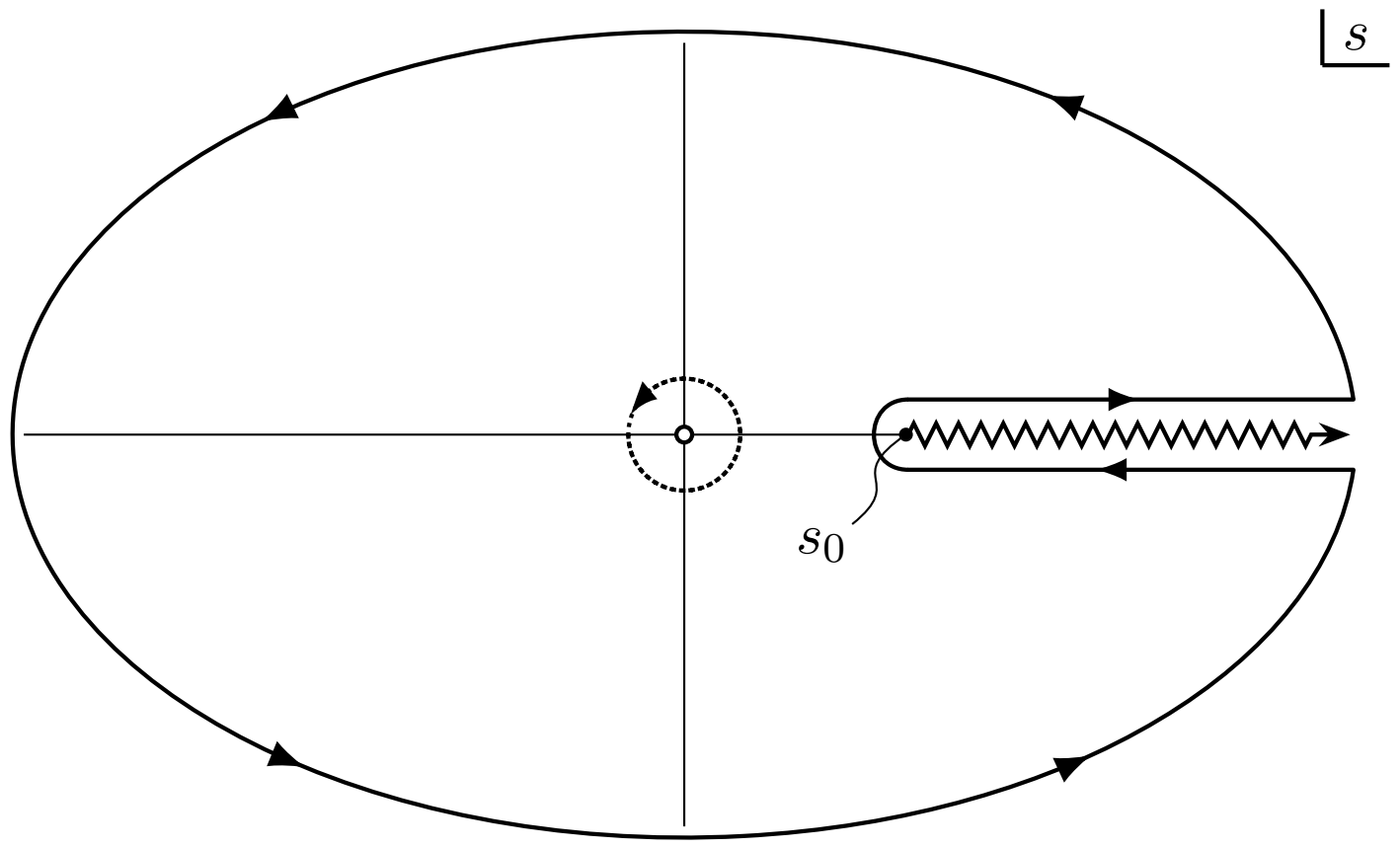
\begin{figure}[t]
    \centering
    \begin{tikzpicture}[arrowmark/.style 2 args={decoration={markings,mark=at position #1 with \arrow{#2}}}, scale=0.75]
        \draw (-5.9,0)-- (2,0);
        \draw (0,-3.5)--(0,3.5);
        \draw[thick] (5.7,3.3)--(6.3,3.3);
        \draw[thick] (5.7,3.3)--(5.7,3.8);
        \draw (6,3.55) node{$s$};
        \draw[thick,fill=white] (0,0) circle(2pt);
        \draw[thick, decoration = {zigzag,segment length = 1.5mm, amplitude = 0.75mm}, decorate] (2,0) -- (5.6,0);
        \draw[thick,-{Stealth[length=6pt]}] (5.6,0)--(5.95,0);
        \filldraw (1.98,0) circle(1.5pt);
        \draw (1.98,0) .. controls (1.4,-0.4) and (2,-0.4) .. (1.5,-0.8) node[below left=-3pt] {$s_0$};
        \draw[thick,densely dotted,-{Latex[length=6pt,bend]}] (-180:0.5) arc[start angle=-180,delta angle=700,radius={0.5}];
        \begin{scope}[yscale=0.6]
            \coordinate (t) at ([shift=(5:6cm)]0,0);
            \coordinate (b) at ([shift=(-5:6cm)]0,0);
            \draw[thick, postaction={decorate},arrowmark={0.135}{Latex[length=7pt,bend]},arrowmark={0.375}{Latex[length=7pt,bend]},arrowmark={0.64}{Latex[length=7pt,bend]},arrowmark={0.88}{Latex [length=7pt,bend]}] ([shift=(5:6cm)]0,0) arc (5:355:6cm);
            \draw[cap=round,thick,postaction={decorate},arrowmark={0.55}{Latex[length=6pt,reversed]}] let \p1 = (t) in (t)--(2,\y1) coordinate (lt);
            \draw[cap=round,thick,postaction={decorate},arrowmark={0.58}{Latex[length=6pt]}] let \p1 = (b) in (b)--(2,\y1) coordinate (lb);
            \draw[thick] (lt) to [out=180, in=180] (lb);
        \end{scope}
    \end{tikzpicture}
    \caption{Integration contours and analytic structure of $\widetilde{G}(s)/s^2$.} 
    \label{fig:TT_comp_s_plane_2d}
\end{figure}

In momentum space, the relation \eqref{TTminusThTh} becomes
\begin{equation}
    \llangle T(p)T(-p)\rrangle - \llangle \opTheta(p)\opTheta(-p)\rrangle = \widetilde{C} p^2\,.
\end{equation}
This is in Euclidean signature. Wick rotating to Lorentzian signature gives, for time-ordered correlators,
\begin{equation}
    \llangle {\cal T}[ T(p)T(-p)]\rrangle - \llangle {\cal T}[ \opTheta(p)\opTheta(-p) ]\rrangle = -i \widetilde{C} p^2\,.
\end{equation}
With $p^2=-s$, we may write
\begin{equation}
    \widetilde{C} = -\frac{1}{2\pi} \oint \frac{ds}{s^2}\lsp \llangle \mathcal{T}[\traceT(p)\traceT(-p)]\rrangle +\frac{1}{2\pi}\beta^I\beta^J\oint \frac{ds}{s^2}\lsp\llangle \mathcal{T}[\O_I (p)\O_J(-p)]\rrangle\,.\label{eq:CtildeSumRule22d}
\end{equation}
The integration contour is a small circle around the origin in the complex $s$-plane. Denote
\begin{equation}
    \widetilde{G}(s) = \llangle \mathcal{T}[\traceT(p)\traceT(-p)]\rrangle\,, \qquad \widetilde{G}_{IJ}(s) = \llangle {\cal T}[ \O_I(p)\O_J(-p)] \rrangle\,.
\end{equation}
The analytic structure of $\widetilde{G}(s)  $ is as shown in Fig.\ \ref{fig:TT_comp_s_plane_2d}. The cut can extend all the way to the origin, but conformal invariance of the IR fixed point means that \eqref{eq:CtildeSumRule22d} nonetheless holds on a vanishingly small circle.

We can evaluate \eqref{eq:CtildeSumRule22d} by steps similar to what we did in position space. The first term is RG-invariant, because $\D\langle TT \rangle =0$. Since the circle is near the origin, we can evaluate this term in the IR CFT and the constant is $C_{\text{IR}}$. The second term can be deformed onto a contour along part of the positive real axis plus a larger circle, so we have
\begin{equation}
    \widetilde{C} = C_{\text{IR}} + \frac{1}{2\pi} \int_0^{\mu^2} \frac{ds}{s^2}\lsp \beta^I \beta^J\Disc \widetilde{G}_{IJ}(s) + \frac{1}{2\pi} \beta^I \beta^J \oint_{|s|=\mu^2} \frac{ds}{s^2}\, \widetilde{G}_{IJ}(s)\, ,
\end{equation}
for any $\mu$. We now choose a renormalization scheme by imposing
\begin{equation}
    \oint_{|s|=\mu^2} \frac{ds}{s^2}\,\widetilde{G}_{IJ}(s) = 0\,.
\end{equation}
This is always possible, because the counterterm in \eqref{t:logZct} contributes $-ic_{IJ} s$ to the correlator and $2\pi c_{IJ}$ to this integral. In this scheme, the $c$-function is
\begin{align}\label{t:pcfunc}
    \widetilde{C} &= C_{\text{IR}} + \frac{1}{2\pi}  \int_0^{\mu^2} \frac{ds}{s^2}\lsp \beta^I \beta^J \Disc \widetilde{G}_{IJ}(s) \\
    &= C_{\text{IR}} + \frac{1}{2\pi} \int_0^{\mu^2} \frac{ds}{s^2} \Disc \widetilde{G}(s)\,,
\end{align}
where in the second line we have used the fact that taking the discontinuity removes contact terms \cite{Komargodski:2011xv}.
When we act on this equation with $\mu \frac{d}{d\mu}$, the only contribution on the right-hand side comes from the endpoint of integration, since $\D\langle TT \rangle = 0$.
Therefore, we obtain the flow equation
\begin{equation}
    \mu \frac{d}{d\mu} \widetilde{C} = \frac{1}{\pi\mu^2} \Disc \widetilde{G}(\mu^2) \geq 0\,.
\end{equation}
The discontinuity is $2\pi \rho(\mu^2)$ with $\rho$ the non-negative spectral function, as reviewed around equation \eqref{oospectral}. We have therefore defined a monotonic $c$-function that depends only on the running couplings. The sum rule is obtained by setting $\lambda = \lambda_{\text{UV}}$, $\mu =\infty$ in \eqref{t:pcfunc},
\begin{equation}
    C_{\text{UV}} - C_{\text{IR}} = \frac{1}{2\pi} \int_0^\infty \frac{ds}{s^2} \Disc \widetilde{G}(s)\,.
\end{equation}

\section{Local RG in four dimensions}\label{s:localrg4d}
In this section we review the local renormalization group in four dimensions, and use it to derive the Callan--Symanzik equations for the ANEC operator.

\subsection{The generalized anomaly}
In four dimensions, the generalized anomaly is \cite{Osborn:1991gm}
\begin{equation}\label{anom4gen}
    \langle \opS\rangle_{g,\lambda} = \langle T - \beta^I\O_I\rangle_{g,\lambda} = {\cal A}_4 = {\cal A}_4^{(1)} + {\cal A}_4^{(2)}\,.
\end{equation}
The first term is
\begin{equation}\label{anom4d1}
    \begin{aligned}
        \mathcal{A}_4^{(1)} &= -a E_4 + \tfrac19 b_1 R^2 + c\lsp W^2\\
        &\quad - \tfrac13 E_I\lsp\partial_\mu \lambda^I \lsp\partial^\mu\logsp R - \tfrac16 F_{IJ}\lsp\partial_\mu \lambda^I \lsp\partial^\mu \lambda^J R - 4H_{IJ}\lsp\partial_\mu \lambda^I \lsp\partial_\nu \lambda^J G^{\mu\nu} \\
        &\quad + \tfrac12 G_{IJ}\lsp\nabla^2 \lambda^I \lsp\nabla^2 \lambda^J - \tfrac12B_{IJK}\lsp\partial_\mu \lambda^I \lsp\partial^\mu \lambda^J \lsp\nabla^2 \lambda^K - \tfrac14C_{IJKL}\lsp\partial_\mu \lambda^I \lsp\partial^\mu \lambda^J \lsp\partial_\nu \lambda^K \lsp\partial^\nu \lambda^L\,,
    \end{aligned}
\end{equation}
with $G_{\mu\nu}= R_{\mu\nu}-\tfrac12 g_{\mu\nu}R$ the Einstein tensor and 
\begin{equation}
    E_4 = R^{\mu\nu\rho\sigma}R_{\mu\nu\rho\sigma}-4 R^{\mu\nu}R_{\mu\nu} + R^2\,,\qquad  
    W^2 = R^{\mu\nu\rho\sigma}R_{\mu\nu\rho\sigma}-2R^{\mu\nu}R_{\mu\nu}+\tfrac13 R^2\,.
\end{equation}
$E_4$ is the usual Euler density in four dimensions and $W^2$ the Weyl tensor squared. The second piece, which does not contribute for constant Weyl variations, is
\begin{equation}
    \begin{aligned}
        \mathcal{A}_4^{(2)}=\nabla^\mu\Big[&G_{\mu\nu}\lsp w_I \lsp\partial^\nu \lambda^I + \tfrac{1}{3}\partial_\mu(b_2 R) + \tfrac{1}{3} R Y_I \lsp\partial_\mu \lambda^I\\
        &+ \partial_\mu\!\left(U_I \lsp\nabla^2 \lambda^I + \tfrac{1}{2} V_{IJ}\lsp\partial_\nu \lambda^I \lsp\partial^\nu \lambda^J\right) + S_{IJ}\lsp\partial_\mu \lambda^I \lsp\nabla^2 \lambda^J + \tfrac{1}{2} T_{IJK}\partial_\nu \lambda^I \lsp\partial^\nu \lambda^J \lsp\partial_\mu \lambda^K\Big]\,.
    \end{aligned}
\end{equation}
All of the coefficients are functions of the spacetime-dependent couplings, and $a(\lambda)$ and $c(\lambda)$ agree with the usual Weyl anomalies at conformal fixed points. We assume a scheme in which the coefficient functions have no explicit dependence on the RG scale $\mu$, but they depend on $\mu$ through the running couplings.

A general local counterterm in four dimensions is
\begin{align}
    \log Z_{\rm ct} &= \int d^{\lsp 4} x\,\sqrt{g}\,\Big[ -\hat a E_4+\tfrac19\hat b R^2+\hat c W^2 -\tfrac13\widehat E_I\,\partial_\mu\lambda^I\partial^\mu R \notag\\
    &\qquad -\tfrac16\widehat F_{IJ}\,\partial_\mu\lambda^I\partial^\mu\lambda^J R -4\widehat H_{IJ}\,\partial_\mu\lambda^I\partial_\nu\lambda^J G^{\mu\nu} +\tfrac12\widehat G_{IJ}\,\nabla^2\lambda^I\nabla^2\lambda^J \notag\\
    &\qquad -\tfrac12\hat b_{IJK}\,\partial_\mu\lambda^I\partial^\mu\lambda^J\nabla^2\lambda^K -\tfrac14\hat c_{IJKL}\,\partial_\mu\lambda^I\partial^\mu\lambda^J \partial_\nu\lambda^K\partial^\nu\lambda^L\Big]\,.
\end{align}
Adding this term shifts the anomaly functions by
\begingroup
\allowdisplaybreaks
\begin{align}
    \delta_{\rm ct}(a,b_1,c,E_I,F_{IJ},H_{IJ},G_{IJ}) &= \mathsterling_{\!\beta}(\hat a,\hat b,\hat c, \widehat E_I,\widehat F_{IJ},\widehat H_{IJ},\widehat G_{IJ})\,,\notag\\
    \delta_{\rm ct}B_{IJK} &= \mathsterling_{\!\beta}\hat b_{IJK} -2\widehat G_{LK}\,\partial_I\partial_J\beta^L\,,\notag\\
    \delta_{\rm ct}C_{IJKL} &= \mathsterling_{\!\beta}\hat c_{IJKL} +\hat b_{KLM}\,\partial_I\partial_J\beta^M +\hat b_{IJM}\,\partial_K\partial_L\beta^M\,,\notag\\
    \delta_{\rm ct}w_I &= -8\partial_I\hat a+8\widehat H_{IJ}\beta^J\,,\notag\\
    \delta_{\rm ct}b_2&=-4\hat b+\widehat E_I\beta^I\,,\label{4dshifts}\\
    \delta_{\rm ct}U_I &= -2\widehat E_I+\widehat G_{IJ}\beta^J\,,\notag\\
    \delta_{\rm ct}Y_I&=-2\widehat E_I -\partial_I(\widehat E_J\beta^J)+\widehat F_{IJ}\beta^J\,,\notag\\
    \delta_{\rm ct}V_{IJ} &= -4\partial_{(I}\widehat E_{J)}+2\widehat F_{IJ} -8\widehat H_{IJ}-\hat b_{IJK}\beta^K\,,\notag\\
    \delta_{\rm ct}S_{IJ} &= 8\widehat H_{IJ}-2\widehat G_{IJ} -2(\partial_I\beta^K)\widehat G_{KJ}+\hat b_{KIJ}\beta^K\,,\notag\\
    \delta_{\rm ct}T_{IJK} &= 16\partial_{(I}\widehat H_{J)K}-8\partial_K\widehat H_{IJ} +2\hat b_{IJK}+2(\partial_K\beta^L)\hat b_{IJL} +2\hat c_{IJKL}\beta^L\,.\notag
\end{align}
\endgroup

\subsection{Weyl consistency conditions}
The anomaly functions are not independent. The Wess--Zumino consistency condition is the vanishing of
\begin{align}\label{eq:wcc4d}
    [\Delta_{\sigma_1}, &\Delta_{\sigma_2}] \log Z = -\Delta_{\sigma_1} \int d^{\lsp 4} x \sqrt{g}\, \sigma_2 {\cal A}_4 +\Delta_{\sigma_2} \int d^{\lsp 4} x \sqrt{g}\, \sigma_1 {\cal A}_4\nonumber\\
    &= \int d^{\lsp 4} x \sqrt{g}\left( 8\p_I a + \beta^J\p_J w_I + w_J \p_I\beta^J - 8H_{IJ}\beta^J\right) (\sigma_2 \partial_\mu \sigma_1 - \sigma_1 \partial_\mu\sigma_2) G^{\mu\nu}\partial_\nu \lambda^I + \cdots\,.
\end{align}
In the second line, we have written only the term proportional to the background Einstein tensor, which imposes the consistency condition
\begin{equation}\label{pIa}
    \p_I \tilde{a} = \left(H_{IJ} + \tfrac14\p_{[I}w_{J]}\right)\beta^J\,, \qquad \tilde{a} = a + \tfrac18 \beta^I w_I\,.
\end{equation}
The other tensor structures in the variation give many other consistency relations that can be found in \cite{Osborn:1991gm}. Contracting \eqref{pIa} with the beta function gives the flow equation
\begin{equation}\label{aflow1}
    \mu \frac{d}{d\mu} \tilde{a} = H_{IJ}\beta^I\beta^J\,.
\end{equation}
This flow equation is clearly analogous to the Zamolodchikov flow equation in two dimensions, but $H_{IJ}$ is not the Zamolodchikov metric. It is by definition the coefficient of $G^{\mu\nu} \partial_\mu \lambda^I \partial_\nu \lambda^J$ in the generalized anomaly. This term has no manifest positivity. The term that is manifestly positive is the 4d Zamolodchikov metric $G_{IJ}$ in \eqref{anom4d1}. At leading order around a perturbative fixed point, these two metrics are in fact proportional. One of the additional Weyl consistency conditions can be expanded to obtain the relation
\begin{equation}
    H_{IJ} = \tfrac14G_{IJ} + O(\beta, \p\beta)\,.
\end{equation}
This establishes the perturbative positivity of $H_{IJ}$, and therefore the perturbative strong $a$-theorem \cite{Osborn:1989td,Osborn:1991gm}. However, it sheds little light on the non-perturbative case.

\subsection{Callan--Symanzik equations}
Following the procedure outlined in section \ref{s:localrgsetup}, we will now vary \eqref{anom4gen} to derive the RG equations for correlation functions. Varying the anomaly relation once gives $\opS$-correlators with a single additional insertion,
\begin{align}
    \llangle \O_I(k) \opS(-k) \rrangle &= -U_I(k^2)^2\,,\\
    \llangle T_{\mu\nu}(k) \opS(-k) \rrangle &= \tfrac{2}{3}b_2 k^2 \left(k_{\mu}k_{\nu}-\delta_{\mu\nu}k^2\right).
\end{align}
At zero momentum these become the RG equations
\begin{align}
    \D \langle \O_I\rangle &= -\p_I\beta^J\,\langle \O_J\rangle\,,\\
    \D\langle T_{\mu\nu}\rangle &= 0\,.
\end{align}
The anomalous dimension term in the first equation comes from \eqref{Dlambdacomm}.

The RG equations of higher-point functions are obtained in the same way: vary the anomaly to calculate the $\opS$-correlator, evaluate at zero momentum, and use \eqref{Dlambdacomm} to translate $\opS(0) \to \D$. The results for 2-point functions are
\begin{align}
    \D\llangle \O_I(k)\O_J(-k)\rrangle &=-\p_I\beta^K\,\llangle\O_K(k)\O_J(-k)\rrangle -\p_J\beta^K\,\llangle\O_I(k)\O_K(-k)\rrangle\notag\\[-5pt]
    &\quad+\p_I\p_J\beta^K\,\langle\O_K\rangle+G_{IJ}(k^2)^2\,,\\
    \D\llangle T_{\mu\nu}(k)\O_I(-k)\rrangle &=-\p_I\beta^J\,\llangle T_{\mu\nu}(k)\O_J(-k)\rrangle-\tfrac23E_I k^2 \left(k_{\mu}k_{\nu}-\delta_{\mu\nu}k^2\right),\\
    \D\llangle T_{\mu\nu}(k)T_{\rho\sigma}(-k)\rrangle &=2c\left(k_{\mu}k_{\rho}-\delta_{\mu\rho}k^2\right) \left(k_{\nu}k_{\sigma}-\delta_{\nu\sigma}k^2\right)+2c\left(k_{\mu}k_{\sigma}-\delta_{\mu\sigma}k^2\right) \left(k_{\nu}k_{\rho}-\delta_{\nu\rho}k^2\right)\notag\\[-5pt]
    &\quad+\tfrac43\left(\tfrac23b_1-c\right) \left(k_{\mu}k_{\nu}-\delta_{\mu\nu}k^2\right) \left(k_{\rho}k_{\sigma}-\delta_{\rho\sigma}k^2\right).
\end{align}
For 3-point functions the full expressions are unwieldy. We will only need traces and the pieces relevant to the ANEC. Define the momentum-space ANEC operator,
\begin{equation}
    E_{uu}(k) = \left. T_{uu}(k)\right|_{k_u = 0}\,.
\end{equation}
This is a Lorentzian momentum, but we can do this replacement formally in Euclidean signature without any ambiguity because the contact terms are analytic so the Wick rotation is just a phase. The RG equations for 3-point functions involving this operator are
\begin{align}\label{DAOO}
    \D\llangle E_{uu}(q) \O_I(k_1) \O_J(k_2)\rrangle &=-\p_I\beta^K\,\llangle E_{uu}(q)\O_K(k_1)\O_J(k_2)\rrangle-\p_J\beta^K\,\llangle E_{uu}(q)\O_I(k_1)\O_K(k_2)\rrangle\notag\\[-5pt]
    &\quad-2k_{1u}^2\left[4H_{IJ}q^2+G_{IJ}(k_1^2+k_2^2)\right],\\
    \D \llangle E_{uu}(q) T(k_1) T(k_2) \rrangle &=-16b_1k_{1u}^2(k_1^2+k_2^2)\,.\label{DATT}
\end{align}
Contracting with beta functions, the first of these equations becomes
\begin{equation}\label{DAThTh}
    \D \llangle E_{uu}(q)\Theta(k_1)\Theta(k_2) \rrangle =-2k_{1u}^2\left[4H_{IJ}q^2+G_{IJ}(k_1^2+k_2^2)\right]\beta^I\beta^J\,.
\end{equation}
These equations are remarkably simple --- the full RG equations  $\D\langle T_{\mu\nu} \O_I \O_J\rangle$ and $\D\langle T_{\mu\nu}TT\rangle$ have many hundreds of terms.

\subsection{The flow equation from correlators}
We will now reproduce the flow equation \eqref{aflow1} for $\tilde{a}$ from the correlation functions and, in the process, see how the combination $\tilde{a} = a + \frac{1}{8}\beta^I w_I$ naturally arises from considering the ANEC operator. Since $T =\beta^I\O_I$ up to contact terms, the difference $
    \langle \E T T \rangle - \langle \E \Theta \Theta \rangle$
must be local. Note that the variations defining $\E$ and $T$ do not commute. With $\E$ between the two traces, the difference would contain the semi-local term $2\llangle T_{uu}(-k_2)T(k_2)\rrangle$. To evaluate the difference, we write
\begin{equation}
    \E T_1T_2 - \E \Theta_1\Theta_2 = \E T_1\opS_2 + \E \opS_1\Theta_2 = \E T_1\opS_2 + \E \Theta_2\opS_1\,,
\end{equation}
where the subscripts label the positions. In the last step we used that the variations defining $\opS$ and $\Theta$ commute. Both terms now have $\opS$ last, so both follow directly from the anomaly. The first term gives $8a\,k_{1u}^2q^2+4b_2k_{1u}^2(k_1^2+k_2^2)$, and the second gives $\beta^Iw_I\,k_{1u}^2q^2+2\beta^IU_I\,k_{1u}^2(k_1^2+k_2^2)$. The resulting identity in momentum space is
\begin{equation}\label{ATTvsAThTh}
    \llangle E_{uu}(q) T(k_1) T(k_2)\rrangle = \llangle E_{uu}(q) \Theta(k_1) \Theta(k_2) \rrangle+8\tilde a\,k_{1u}^2q^2
    +2\left(2b_2+\beta^IU_I\right)k_{1u}^2(k_1^2+k_2^2)\,.
\end{equation}
Let's collect the essential results so far, making one further simplification which is to insert the traces at null momenta, $k_1^2 = k_2^2 = 0$. In these kinematics, equations \eqref{DATT}, \eqref{DAThTh}, and \eqref{ATTvsAThTh} become
\begin{align}
    \D\llangle E_{uu}(q) T(k_1) T(k_2) \rrangle &= 0\,,\\
    \D\llangle E_{uu}(q) \Theta(k_1) \Theta(k_2)\rrangle &= -8 k_{1u}^2 q^2 H_{IJ}\beta^I \beta^J\,,\\
    \llangle E_{uu}(q) T(k_1) T(k_2)\rrangle &= \llangle E_{uu}(q) \Theta(k_1) \Theta(k_2) \rrangle+8\tilde a\,k_{1u}^2q^2\,.
\end{align}
Taking the $\D$ derivative of the last equation and using the first two gives
\begin{equation}
    0 = 8\left(-H_{IJ}\beta^I\beta^J +\D\tilde{a}\right)q^2 k_{1u}^2\,.
\end{equation}
Thus, we have reproduced the flow equation \eqref{aflow1} for $\tilde{a}$. In terms of correlation functions, this flow equation is simply the RG equation for $\langle T \E T \rangle$.

The uncontracted flow equation \eqref{pIa} can be reproduced in the same way. At null momenta, the relevant RG equations and 3-point identity are
\begin{align}
    \D \llangle E_{uu}(q) T(k_1) \O_I(k_2)\rrangle &= -\p_I\beta^J\lsp\llangle E_{uu}(q)T(k_1)\O_J(k_2)\rrangle -8\p_Ia\,k_{1u}^2q^2\,,\\
    \D\llangle E_{uu}(q) \Theta(k_1) \O_I(k_2) \rrangle &= -\p_I\beta^J\lsp\llangle E_{uu}(q)\Theta(k_1)\O_J(k_2)\rrangle -8H_{IJ}\beta^J k_{1u}^2q^2\,,\\
    \llangle E_{uu}(q) T(k_1) \O_I(k_2) \rrangle &= \llangle E_{uu}(q) \Theta(k_1) \O_I(k_2)\rrangle +w_I k_{1u}^2q^2\,.
\end{align}
Acting with $\D$ on the last equation gives \eqref{pIa}.

\subsection{Integral relations}
As in two dimensions, we can now invert the flow equations to solve for the anomaly functions as integrated correlation functions. We work in Lorentzian signature so that we can use the ANEC operator. All of our Euclidean results continue trivially to time-ordered correlators in Lorentzian signature, with each $n$-point function picking up a phase of $(-i)^{n-1}$ as explained in appendix \ref{app:corr}. Let's start with $\tilde{a}$, which we will extract from \eqref{ATTvsAThTh}. Taking derivatives with respect to momenta and then setting the momenta to zero we obtain
\begin{align}\label{afunctionI}
    \tilde{a} &= \frac{1}{64} \int d^{\lsp 4} x_1 d^{\lsp 4} x_2 (u_1-u_2)^2 \vec{y}_1 \cdot \vec{y}_2 \left( -\langle {\cal T}[T_{uu}(0) T(x_1)T(x_2) ]\rangle + \langle {\cal T}[ T_{uu}(0) \Theta(x_1)\Theta(x_2)]\rangle \right)\\
    &= \frac{1}{64} \int d^{\lsp 4} x_1 d^{\lsp 4} x_2 (u_1-u_2)^2 \vec{y}_1 \cdot \vec{y}_2 \left( -\langle {\cal T}[T_{uu}(0) T(x_1)T(x_2) ]\rangle + \beta^I\beta^J \langle {\cal T}[ T_{uu}(0) \O_I(x_1)\O_J(x_2)]\rangle \right),\notag
\end{align}
where $\vec{y}_i$ are the transverse positions. The contact terms added to the integrand in the second line drop out upon integration. We have used translations to fix the null energy insertion at the origin.

The analogous relation for the Jack--Osborn metric, from inverting \eqref{DAOO}, is
\begin{equation}\label{HIJinversion}
    H_{IJ} = \frac{1}{64}\int d^{\lsp 4}x_1\,d^{\lsp 4}x_2\,(u_1-u_2)^2\lsp \vec{y}_1\cdot \vec{y}_2 \lsp \mathsterling_{\!\beta} \langle {\cal T}[T_{uu}(0)\O_I(x_1)\O_J(x_2)]\rangle\,.
\end{equation}
Both of these results are closely analogous to the 2d results for $\widetilde{C}$ and the metric $\chi_{IJ}$ derived in section \ref{ss:2dintegralrelations}, but with an additional insertion of the ANEC operator and a different integration kernel. In \eqref{HIJinversion} we dropped a term with $\mu \frac{\p}{\p \mu}$, which is justified by the argument under \eqref{eq:chiIJasLie}.

\subsection{The ANEC sum rule}
The function $\tilde{a}$ in \eqref{afunctionI} depends on the running couplings. By evaluating it at the UV and IR fixed points we can reproduce the ANEC sum rule for $\Delta a$ \cite{Hartman:2023qdn}.

The first term in \eqref{afunctionI} is independent of $\mu$ by \eqref{DATT}. The second term vanishes at an IR fixed point where $\beta^I=0$ and the integral is finite. At a UV fixed point, the second term contributes  away from coincident points only, by the same logic as in section \ref{ss:JOcsumrule}. Therefore,
\begin{equation}
    \Delta a = \frac{1}{64} \int d^{\lsp 4} x_1 d^{\lsp 4} x_2 (u_1-u_2)^2 \vec{y}_1 \cdot \vec{y}_2 \langle {\cal T}[ T_{uu}(0) \O_I(x_1) \O_J(x_2) ]\rangle_{\rm sep}\lsp \beta^I \beta^J\,,
\end{equation}
where `sep' indicates that contact terms are dropped. This is exactly the sum rule obtained in \cite{Hartman:2023qdn}. Ordinarily one cannot uniquely divide a distribution into its contact part and separated part, but this ambiguity drops out of the integral as discussed in detail in \cite{Hartman:2023qdn}. In any case, we will now apply the spectral analysis from sections \ref{s:spectral}--\ref{s:exacta} to give a cleaner derivation that avoids separating the correlator in this way, and moreover defines the monotonic $a$-function from the ANEC.\footnote{There is also an ANEC sum rule for $\Delta c$ in two dimensions \cite{Hartman:2023ccw}, and a corresponding relation between the two-dimensional Jack--Osborn metric and the ANEC-type metric. However, in two dimensions the ANEC metric is essentially the same as the Zamolodchikov metric, because $\E_u = - P_u$ simply acts as a derivative. }

\section{The Jack--Osborn metric and the ANEC}\label{s:jometric}

We will now apply the exact framework of sections \ref{s:spectral} and \ref{s:exacta} to the four-dimensional perturbative RG flows studied in section \ref{s:localrg4d}. We will repeat some of the key formulas to make this section more readable.  

The coefficients in the generalized anomaly ${\cal A}_4$ are scheme-dependent, with the transformations \eqref{4dshifts} under adding local counterterms. The flow equation \eqref{aflow1} holds in any scheme, but the function $\tilde{a}(\lambda)$ along the flow and the metric $H_{IJ}$ are not invariant. Hence, the question of positivity and monotonicity is whether there exists a scheme in which $H_{IJ} \succeq 0$.\footnote{There are certainly schemes where $H_{IJ}$ is negative or indefinite. An example is minimal subtraction in large-$N_f$ gauge theories \cite{Antipin:2018brm}.} Non-perturbatively, we derived a positive metric in section \ref{s:exacta}, proportional to the ANEC spectral density, and we will now show that in perturbative RG flows there is an explicit scheme in which the Jack--Osborn metric $H_{IJ}$ agrees with the non-perturbative one. The analogous calculation in two dimensions was done in section \ref{s:czamo}, where we described the scheme in which the flow of the $c$-function is dictated by the positive-definite Zamolodchikov metric. 

In a theory deformed by marginally relevant operators $\O_I$, denote the time-ordered correlation functions with the ANEC by
\begin{equation}
    \GE_{IJ}(t) = \omega^{-2}\llangle {\cal T}[E_{uu}(q)\O_I(k_1)\O_J(k_2)]\rrangle\,,
\end{equation}
where we have specialized the kinematics to
\begin{equation}
    k_1^2 = k_2^2 = 0\,, \qquad t = -(k_1+k_2)^2\,, \qquad k_{1u} = -k_{2u} = \omega > 0\,.
\end{equation}
In the same kinematics,
\begin{equation}
    \GE_{TT}(t) = \omega^{-2}\llangle {\cal T}[E_{uu}(q)T(k_1)T(k_2)]\rrangle\,,
\end{equation}
and we use the same notation $\GE_{\Theta\Theta}$ for the correlator with $\Theta = \beta^I\O_I$. In impact parameter space, as in section \ref{s:spectral}, we denote these correlators by
\begin{equation}\label{eq:impparcorrs}
    \FE_{IJ}(b) = \int \frac{d^{\lsp 2}\vec{q}}{(2\pi)^2}e^{-i\vec{q}\cdot \vec{b}}\GE_{IJ}(t=-\vec{q}^{\,2})\,, \qquad \FE_{TT}(b) = \int \frac{d^{\lsp 2}\vec{q}}{(2\pi)^2}e^{-i\vec{q}\cdot \vec{b}}\GE_{TT}(t=-\vec{q}^{\,2})\,.
\end{equation}
One of the main results of the local RG analysis was \eqref{ATTvsAThTh}, which in this notation reads
\begin{equation}
    \GE_{TT}(t) = \GE_{\Theta\Theta}(t) + 8 \tilde{a} t\,,
\end{equation}
where $\tilde{a} = \tilde{a}(\lambda)$ is the $a$-function along the RG flow. Therefore,
\begin{equation}
    \tilde{a} = \tfrac18 \left. \p_{t}\right|_{t=0} \left[ \GE_{TT}(t) - \GE_{\Theta\Theta}(t) \right].
\end{equation}
Transforming to impact parameter space, the derivative $\p_t$ becomes an insertion of $\vec{b}^{\,2}$, and we obtain
\begin{equation}
    \tilde{a} = \frac{1}{32} \int d^{\lsp 2} \vec{b} \, \vec{b}^{\,2} \left[ \FE_{TT}(b) - \FE_{\Theta\Theta}(b) \right] = a_{\text{IR}} - \frac{1}{32} \int d^{\lsp 2} \vec{b} \, \vec{b}^{\,2} \FE_{IJ}(b)\beta^I\beta^J\,,\label{aF1}
\end{equation}
where in the second expression we used the fact \eqref{DATT} that the first term is $\mu$-independent, and evaluated the constant at the IR fixed point. All of these expressions include contact terms.

The relation \eqref{aF1} is scheme-dependent. It is the 4d analog of the Jack--Osborn $c$-function in the form \eqref{JOcBeforeScheme}, before choosing the Zamolodchikov scheme that gave a monotonic flow. Guided by the analysis in two dimensions, we now impose the renormalization condition\footnote{Another natural choice would be to introduce a second scale, $\Lambda$, and cut off the integral at $|\vec{b}| = \Lambda^{-1}$. This leads to an $a$-function with explicit scale dependence in addition to the implicit scale dependence through the couplings: $\tilde{a} = \tilde{a}(\frac{\mu}{\Lambda}, \lambda(\mu))$. This is perfectly consistent (and used for example in \cite{Friedan:2009ik}), but we have assumed a scheme with no explicit $\mu$-dependence in our treatment of the local RG, so some of the equations above would need to be modified. For example, we used $\D \tilde{a} = \beta^I \p_I \tilde{a}$, which does not hold in the scheme with cutoff  $\Lambda^{-1}$.}
\begin{equation}\label{rgschemeF}
    \int_{|\vec{b}|<\mu^{-1}} d^{\lsp 2} \vec{b} \, \vec{b}^{\,2} \FE_{IJ}(b) = 0\,.
\end{equation}
This is always possible. There is a local counterterm
\begin{equation}
    \log Z_{\rm ct} \supset -4\int d^{\lsp 4} x \sqrt{g}\, \widehat{H}_{IJ} \p_\mu \lambda^I \p_\nu \lambda^J G^{\mu\nu}
\end{equation}
that shifts $H_{IJ} \to H_{IJ} + \mathsterling_{\!\beta} \widehat{H}_{IJ}$. The time-ordered correlator $\GE_{IJ}(t)$ shifts by $-8\widehat{H}_{IJ} t$, and the integral in \eqref{rgschemeF} shifts by $-32 \widehat{H}_{IJ}$, allowing us to set it to zero.

In this scheme, the $a$-function \eqref{aF1} is
\begin{equation}
    \tilde{a} = a_{\text{IR}} - \frac{1}{32} \int_{|\vec{b}|>\mu^{-1}}d^{\lsp 2}\vec{b} \, \vec{b}^{\,2} \FE_{IJ}(b)\beta^I\beta^J\,.
\end{equation}
Finally, we use the result of section \ref{ss:relationTO}, where we showed that for $\vec{b}\neq 0$,
\begin{equation}
    \FE_{IJ}(b) = -\rhoE_{IJ}(b)\,,
\end{equation}
where $\rhoE_{IJ}(b) \succeq 0$ is the positive semidefinite spectral density. Thus, we have
\begin{equation}\label{afinalP}
    \tilde{a} = a_{\text{IR}} + \frac{1}{32}\int_{|\vec{b}|>\mu^{-1}} d^{\lsp 2}\vec{b} \, \vec{b}^{\,2} \rhoE_{IJ}(b)\beta^I \beta^J\,.
\end{equation}
This matches the non-perturbative result \eqref{npafunction}, but the interpretation is slightly different. In the non-perturbative result we introduced $\mu$ by hand as a cutoff on the integral, and the $a$-function was $\tilde{a} = \tilde{a}(\mu)$. In \eqref{afinalP}, this is the $a$-function of the local RG, in the scheme \eqref{rgschemeF}, and as such is a function of the running couplings, $\tilde{a} = \tilde{a}(\lambda)$. Its dependence on the RG scale enters entirely through the running.

By acting on \eqref{afinalP} with $\mu \frac{d}{d\mu}$ and comparing to the Jack--Osborn metric, we obtain
\begin{equation}
    H_{IJ}\beta^I \beta^J = \frac{\pi}{16 \mu^4}\rhoE_{IJ}(\mu^{-1})\beta^I\beta^J\,.
\end{equation}
In fact the stronger condition holds without the beta-function contractions. To derive it, we use \eqref{DAOO}, which for the current kinematics reads
\begin{equation}
    \D\GE_{IJ}(t) + \p_I\beta^K \GE_{KJ}(t) + \p_J\beta^K \GE_{IK}(t)= -8H_{IJ} t\,.
\end{equation}
In impact parameter space,
\begin{equation}\label{DFbb}
    \D \FE_{IJ}(b) + \p_I \beta^K \FE_{KJ} + \p_J \beta^K \FE_{IK} = -8 H_{IJ} \p_{\vec{b}}^2 \delta^{(2)}(\vec{b})\,.
\end{equation}
Now we act on the renormalization condition \eqref{rgschemeF} with $\D$, demanding this condition is preserved. This requires
\begin{align}
    0 &= \int_{|\vec{b}| < \mu^{-1}} d^{\lsp 2}\vec{b}\,\vec{b}^{\,2} \D \FE_{IJ}(b) - \frac{2\pi}{\mu^4}\FE_{IJ}(\mu^{-1}) \\
    &= -32 H_{IJ} + \frac{2\pi}{\mu^4} \rhoE_{IJ}(\mu^{-1})\,.
\end{align}
In the second line we used \eqref{DFbb}; the anomalous dimension terms drop out because they multiply the original renormalization condition. Thus, we conclude
\begin{equation}\label{eq:metscheme4d}
    H_{IJ} = \frac{\pi}{16\mu^4}\rhoE_{IJ}(\mu^{-1})\,.
\end{equation}
The Jack--Osborn metric in this scheme is the positive semidefinite ANEC spectral density.

\section{Comments on gradient RG flow}\label{s:gradient}
If a scheme existed in which the anomaly $w_I$ were an exact form, then \eqref{pIa} would become
\begin{equation}
    \partial_I\tilde{a}=H_{IJ}\beta^J\,,
\end{equation} 
and if $H_{IJ}$ were a positive-definite metric in the same scheme then the RG flow would be gradient.\footnote{In two dimensions, the gradient question has been discussed at leading order in conformal perturbation theory in \cite{Zamolodchikov:1986gt}, and in \cite{Gomis:2015yaa} on conformal manifolds, with the conclusion that $w_I$ is exact in those cases. A non-perturbative discussion of the gradient formula was given in \cite{Friedan:2009ik}. For an example where the gradient property might not be valid in $d=2$ see \cite{Freedman:2005wx}.} In perturbation theory in $d=4$, the gradient property holds to high loop order for scalar and scalar-fermion systems \cite{Jack:2018oec,Pannell:2024sia,Pannell:2025ixz,Pannell:2025ajf}, due to non-trivial constraints among beta function coefficients --- see \cite{Stergiou:phi4GraphAtlas} for results to seven loops in multiscalar systems.

Within our framework,
\begin{equation}\label{ATOvsAThO}
    \llangle E_{uu}(q) T(k_1) \O_I(k_2)\rrangle = \llangle E_{uu}(q) \Theta(k_1) \O_I(k_2) \rrangle + w_Ik_{1u}^2q^2 + 2U_Ik_{1u}^2(k_1^2+k_2^2)\,.
\end{equation}
The derivatives that gave \eqref{afunctionI} annihilate the $U_I$ term, as they did the last term in \eqref{ATTvsAThTh}, and we obtain
\begin{align}\label{wIintegral}
    w_I &= \frac{1}{8} \int d^{\lsp 4} x_1 d^{\lsp 4} x_2 (u_1-u_2)^2 \vec{y}_1 \cdot \vec{y}_2 \left( -\langle {\cal T}[T_{uu}(0) T(x_1)\O_I(x_2) ]\rangle + \langle {\cal T}[ T_{uu}(0) \Theta(x_1)\O_I(x_2)]\rangle \right)\notag\\
    &= \frac{1}{8} \int d^{\lsp 4} x_1 d^{\lsp 4} x_2 (u_1-u_2)^2 \vec{y}_1 \cdot \vec{y}_2 \left( -\langle {\cal T}[T_{uu}(0) T(x_1)\O_I(x_2) ]\rangle + \beta^J \langle {\cal T}[ T_{uu}(0) \O_J(x_1)\O_I(x_2)]\rangle \right).
\end{align}
This is the 4d analog of the 2d formula \eqref{eq:2dwformula} for $w_I$. With the ANEC, a single 3-point function suffices.

To connect to the scheme we used in the previous section, define
\begin{equation}
    \GE_{TI}(t) = \omega^{-2}\llangle {\cal T}[E_{uu}(q)T(k_1)\O_I(k_2)]\rrangle\,,
\end{equation}
and let $\FE_{TI}(b)$ be its transverse Fourier transform. In this notation, \eqref{ATOvsAThO} reads $\GE_{TI}(t) = \beta^J\GE_{JI}(t) + w_I t$. In impact parameter space,
\begin{equation}
    \FE_{TI}(b) = \beta^J\FE_{JI}(b) + w_I\lsp \p_{\vec{b}}^2\delta^{(2)}(\vec{b})\,.
\end{equation}
Multiplying by $\vec{b}^{\,2}$ and integrating over $|\vec{b}|<\mu^{-1}$ we find, after using the condition \eqref{rgschemeF},
\begin{equation}\label{wANEC}
    w_I = \frac{1}{4}\int_{|\vec{b}|<\mu^{-1}} d^{\lsp 2}\vec{b}\,\vec{b}^{\,2}\,\FE_{TI}(b)\,.
\end{equation}
The integral includes the contact terms at $\vec{b}=0$. As a check, let us act on \eqref{wANEC} with $\D$. Using $\D\GE_{TI} + \p_I\beta^J\GE_{TJ} = -8\p_I a\, t$ from section \ref{s:localrg4d}, and $\FE_{TI} = -\rhoE_{IJ}\beta^J$ at $|\vec{b}| = \mu^{-1}$, we obtain
\begin{equation}
    \beta^J\p_J w_I + w_J\p_I\beta^J + 8\p_I a = \frac{\pi}{2\mu^4}\rhoE_{IJ}(\mu^{-1})\beta^J\,.
\end{equation}
This is the Weyl consistency condition behind \eqref{pIa}, with $H_{IJ}$ as found in \eqref{eq:metscheme4d}.

The two-form $\Omega_{IJ} = \p_I w_J - \p_J w_I$ is unambiguous in the scheme that gives \eqref{wANEC}, for under a change of local counterterms $\GE_{TI}$ shifts only through the Euler term by $-8\p_I\hat a\, t$. Thus, $\Omega_{IJ}$ cannot be removed and \eqref{pIa} becomes
\begin{equation}\label{pIaANEC}
    \p_I\tilde{a} = \frac{\pi}{16\mu^4}\rhoE_{IJ}(\mu^{-1})\beta^J + \frac{1}{8}\Omega_{IJ}\beta^J\,.
\end{equation}
Requiring $\Omega_{IJ}\beta^J = 0$ needs the one-form $\rhoE_{IJ}(\mu^{-1})\beta^J$ to be closed,
\begin{equation}\label{rhoclosed}
    \p_I\big[\rhoE_{JK}(\mu^{-1})\beta^K\big] = \p_J\big[\rhoE_{IK}(\mu^{-1})\beta^K\big]\,.
\end{equation}
Positivity of $\rhoE_{IJ}$ does not imply \eqref{rhoclosed}.\footnote{In fact, we have only shown that $\rhoE_{IJ}$ is positive semidefinite, not positive definite, which makes the gradient property even harder to prove with our methods. Perhaps the methods of \cite{OpenAI_SvC} can be used to show that the metric is strictly positive definite.} One may relate $\Omega_{IJ}$ to the 4-point function $\langle E_{uu} T \O_J \O_I\rangle$ with $\O_I$ at zero momentum, minus the same with $I$ and $J$ exchanged, using the fact that a coupling derivative inserts an integrated marginal operator at zero momentum. However, exchange symmetry of the 4-point function does not make this vanish, because it also exchanges the momenta of the two operators. Therefore, in the scheme in which the Jack--Osborn metric is given by $\rhoE_{IJ}$, the RG flow is not obviously gradient.

\begin{acknowledgements}
    We thank Shounak De, Zohar Komargodski, Petr Kravchuk, Juan Maldacena, Ian Moult, Hugh Osborn, Julio Parra-Martinez and Jo\~ao Penedones for discussions. 
    
    GM is supported by the Swiss National Science Foundation through the project 200020 197160 and through the National Centre of Competence in Research SwissMAP. AS is supported by the Royal Society under grant URF{\textbackslash}R1{\textbackslash}211417 and by STFC under grant ST/X000753/1. This work was performed in part at Aspen Center for Physics, which is supported by National Science Foundation grant PHY-2210452 and by a grant from the Simons Foundation (1161654, Troyer).  
\end{acknowledgements}

\appendix

\section{Definition of correlation functions}\label{app:corr}

\subsection{Conventions}
In the Euclidean theory, with path integral weighted by $e^{-S}$, the classical stress tensor and marginal operators are defined by $T_{\mu\nu} = \frac{2}{\sqrt{g}} \frac{\delta}{\delta g^{\mu\nu}}S$, $\O_I = \frac{1}{\sqrt{g}} \frac{\delta S}{\delta \lambda^I}$. Define $\opS = T - \opTheta$ with $\opTheta = \beta^I \O_I$. Connected correlation functions are defined by varying $\log Z$, with the convention that $\sqrt{g}$'s go outside the variation but metric contractions and $\beta^I$ contractions go inside:

\begin{align}
    \langle T_{\mu\nu}(x_1) \dots T_{\alpha\beta}(x_n) \rangle &= \frac{(-2)^n}{\prod_{a=1}^n \sqrt{g(x_a)}} \frac{\delta^n \log Z}{\delta g^{\mu\nu}(x_1)\cdots\delta g^{\alpha\beta}(x_n)}\,,\\
    \langle T(x_1) \dots T(x_n) \rangle &= \frac{(-2)^n}{\prod_{a=1}^n \sqrt{g(x_a)}} \left(g^{\mu_1\nu_1}(x_1)\frac{\delta}{\delta g^{\mu_1\nu_1}(x_1)}\right)\cdots \left(g^{\mu_n\nu_n}(x_n)\frac{\delta}{\delta g^{\mu_n\nu_n}(x_n)}\right)\log Z\,,\notag\\
    \langle \O_{I_1}(x_1) \dots \O_{I_n}(x_n) \rangle &= \frac{(-1)^n}{\prod_{a=1}^n \sqrt{g(x_a)}} \frac{\delta^n \log Z}{\delta\lambda^{I_1}(x_1)\cdots\delta\lambda^{I_n}(x_n)}\,,\notag\\
    \langle \opTheta(x_1) \dots \opTheta(x_n)\rangle &= \frac{(-1)^n}{\prod_{a=1}^n \sqrt{g(x_a)}} \left(\beta^{I_1}(\lambda(x_1))\frac{\delta}{\delta\lambda^{I_1}(x_1)}\right)\cdots \left(\beta^{I_n}(\lambda(x_n))\frac{\delta}{\delta\lambda^{I_n}(x_n)}\right)\log Z\,,\notag\\
    \langle \opS(x_1) \dots \opS(x_n) \rangle &= \frac{(-1)^n}{\prod_{a=1}^n \sqrt{g(x_a)}} \prod_{a=1}^n \left(2g^{\mu_a\nu_a}(x_a)\frac{\delta}{\delta g^{\mu_a\nu_a}(x_a)} -\beta^{I_a}(\lambda(x_a))\frac{\delta}{\delta\lambda^{I_a}(x_a)}\right)\log Z\,.\notag
\end{align}

With this convention, $[T(x), T(y)] = [\opS(x),\opS(y)] = [\opTheta(x),\opTheta(y)] = 0$ in Euclidean signature, but $[T_{\mu\nu}(x),T(y)]$ and $[\O_I(x),\opTheta(y)]$ are non-vanishing at coincident Euclidean points. In mixed Euclidean correlators, variations are done in the order written, for example
\begin{equation}
    \langle T_{\mu\nu}(x_1) T(x_2) \rangle = \frac{4}{\sqrt{g(x_1)}\sqrt{g(x_2)}} \frac{\delta}{\delta g^{\mu\nu}(x_1)}\,g^{\alpha\beta}(x_2) \frac{\delta}{\delta g^{\alpha\beta}(x_2)}\log Z\,.
\end{equation}
We have therefore
\begin{align}
    \langle T_{\mu\nu}(x)T(0)\rangle &= \delta^{\alpha\beta}\langle T_{\mu\nu}(x)T_{\alpha\beta}(0)\rangle - 2\langle T_{\mu\nu}(0)\rangle\,\delta^{(d)}(x)\,,\\
    \langle \O_I(x)\opTheta(0)\rangle &= \beta^J \langle \O_I(x)\O_J(0)\rangle - \partial_I\beta^J\,\langle \O_J(0)\rangle\,\delta^{(d)}(x)\,.
\end{align}
We use
\begin{equation}
    \frac{\delta g^{\mu\nu}(x)}{\delta g^{\rho\sigma}(y)} = \delta^{\mu}_{(\rho}\delta^{\nu}_{\sigma)}\delta^{(d)}(x,y)\,, \qquad \frac{\delta\lambda^I(x)}{\delta \lambda^J(y)} = \delta^I_J\lsp \delta^{(d)}(x,y)\,,\label{eq:Guvvardef}
\end{equation}
where $\delta^{(d)}(x,y)$ is the $d$-dimensional Dirac density and $\delta_{\mu\nu}$ the flat Euclidean metric. The Dirac density is related to the standard scalar delta function $\delta^{(d)}(x-y) $ through $\sqrt{g}\,\delta^{(d)}(x-y) = \delta^{(d)}(x,y)$, and $\int d^{\lsp d}x\sqrt{g}\,\delta^{(d)}(x) = 1$. Our curvature conventions are $R^\rho{}_{\!\sigma\mu\nu} = \partial_\mu\Gamma^\rho_{\nu\sigma}-\partial_\nu\Gamma^\rho_{\mu\sigma}+\Gamma^\rho_{\mu\lambda}\Gamma^\lambda_{\nu\sigma}-\Gamma^\rho_{\nu\lambda}\Gamma^\lambda_{\mu\sigma}$ and $R_{\mu\nu} = R^\rho{}_{\!\mu\rho\nu}$, so that spheres have $R>0$.

Our Fourier transform convention is
\begin{equation}
    \hat{f}(k) = \int d^{\lsp d} x\, e^{ik\cdot x}f(x)\,,
\end{equation}
and momentum-space correlators with the delta function removed are denoted by double brackets:
\begin{equation}
    \langle \O(k_1)\dots \O(k_n) \rangle = (2\pi)^d \delta^{(d)}(k_1+\cdots+k_n) \llangle \O(k_1)\dots \O(k_n) \rrangle\,.\label{eq:DoubleBrackets}
\end{equation}

\subsection{Lorentzian correlators}
In Lorentzian signature, the position-space Wightman functions are defined by analytically continuing the Euclidean correlators, with no additional factors. With these conventions, which differ from much of the literature on 2d CFT, the averaged null energy condition is $\int du\, T_{uu} \geq 0$ with $u$ a forward-pointing affine null coordinate.

In momentum space, the correlators pick up an additional phase under Wick rotation. Denote a Euclidean $n$-point correlator by
\begin{equation}
    G_E(p_1,\dots,p_{n-1}) = \llangle \O(p_1) \cdots \O(p_{n-1}) \O(-p_1-\dots-p_{n-1}) \rrangle\,,
\end{equation}
and a time-ordered Lorentzian correlator by
\begin{equation}
    G_T(p_1,\dots,p_{n-1}) = \llangle {\cal T}[ \O(p_1) \cdots \O(p_{n-1})\O(-p_1-\dots-p_{n-1}) ]\rrangle\,.
\end{equation}
The relation, for real Minkowski momenta $p_i$, is
\begin{equation}
    G_T(p_1,\dots,p_{n-1}) = (-i)^{n-1} G_E(p_1, \dots, p_{n-1})\,,
\end{equation}
where the right-hand side is defined by analytic continuation along the usual contour: $p_i^0(\theta) = e^{i\theta}p_i^0$, starting at $\theta = \frac{\pi}{2}$ (Euclidean) and decreasing to $\theta  = 0$ (Minkowski), with spatial momenta held fixed. The phase comes from the measure in
\begin{equation}
    G_E(k_1,\dots,k_{n-1}) = \int \prod_{a=1}^{n-1} d^{\lsp d} x_a\, e^{i(k_1\cdot x_1+\dots + k_{n-1}\cdot x_{n-1})} \langle \O(x_1) \dots \O(0) \rangle\,.
\end{equation}

\section{Momentum spectral function}\label{app:momentum}

In this appendix we briefly consider the spectral representation of $\langle \O \E \O \rangle$ in momentum space rather than impact parameter space. This appendix is not used in the applications in the main text, but there is a nice relation to the double discontinuity that could be useful elsewhere.

The construction is analogous to \eqref{oospectral}, except that $P^2$ does not commute
with $\E$, so the $\delta(P^2+s)$ projector must be inserted twice. We define the spectral function by
\begin{equation}\label{momentumspectral}
    \rhoE_{\O\O}(s_1, s_2; t) = (2\pi)^5 \langle \O| \delta^{(2)}(\vec{P} - \vec{q}_1) \delta(P^2+s_1) \E \delta(P^2+s_2) \delta^{(2)}(\vec{P} - \vec{q}_2) \delta(P_u+\omega) |\O\rangle\,,
\end{equation}
with $|\O\rangle = \O(x=0)|0\rangle$ and $t=-(\vec{q}_1-\vec{q}_2)^2$. The ANEC, $\E\succeq 0$, implies that this is a positive semidefinite kernel, meaning 
\begin{equation}
    \int ds_1\, ds_2\, d^{\lsp 2} \vec{q}_1\, d^{\lsp 2} \vec{q}_2 \,f^*(s_1,\vec{q}_1) \rhoE_{\O\O}(s_1,s_2;t) f(s_2,\vec{q}_2) \geq 0\,.
\end{equation}
The spectral function in impact parameter space is the transverse Fourier transform,
\begin{equation}
    \rhoE_{\O\O}(s_1,s_2;b) = \int \frac{d^{\lsp 2} \vec{q}}{(2\pi)^2} e^{-i\vec{q}\cdot \vec{b}} \rhoE_{\O\O}(s_1,s_2;t=-\vec{q}^{\,2})\,.
\end{equation}
The spectral function \eqref{momentumspectral} is the double discontinuity of the time-ordered correlator,
\begin{equation}\label{dDiscG}
    \dDisc \GE_{\O\O}(s_1,s_2; t) = 16\pi^2 \rhoE_{\O\O}(s_1,s_2;t)\,,
\end{equation}
with the double discontinuity taken with respect to $s_1$ and $s_2$ at fixed $t$,
\begin{align}\label{eq:dDiscdef}
    \dDisc G(s_1,s_2; t) &:= \Disc_{s_1}\Disc_{s_2} G(s_1,s_2; t) \\
    &= G(s_1+i\epsilon,s_2+i\epsilon;t) + G(s_1-i\epsilon,s_2-i\epsilon;t)\notag\\
    &\quad - G(s_1-i\epsilon,s_2+i\epsilon;t) - G(s_1+i\epsilon,s_2-i\epsilon;t)\,.\notag
\end{align}
To derive this, start from \eqref{towight},
\begin{equation}
    \GE_{\O\O}(s_1,s_2;t) = \frac{1}{\omega^2} \int d^{\lsp 4} x_1\, d^{\lsp 4} x_2\, e^{ik_1\cdot x_1+ik_2\cdot x_2}\theta(v_1)\theta(-v_2) \langle \O(x_1)T_{uu}(0) \O(x_2) \rangle + \text{contact}\,.
\end{equation}
Write the correlator in the integrand as $\langle \O| e^{i P\cdot x_1} T_{uu}(0) e^{-i P \cdot x_2}|\O\rangle$ and do the $x$ integrals,  using that the half-line integral is 
\begin{equation}
    \int_0^\infty dv \,e^{i\kappa v} =\frac{i}{\kappa+i\epsilon}\, ,\label{eq:halfintegral} 
\end{equation}
then use
\begin{equation}
    \delta(P_u+\omega)T_{uu}(0)\delta(P_u+\omega) =\frac{1}{2\pi}\E \delta(P_u+\omega)\,.
\end{equation}
This yields
\begin{align}\label{gaad}
    &\GE_{\O\O}(s_1,s_2;t)=\\
    &\qquad -4(2\pi)^5 \langle \O |\delta^{(2)}(\vec{P} - \vec{q}_1) \frac{1}{P^2+s_1+i\epsilon} \E \frac{1}{P^2+s_2+i\epsilon} \delta^{(2)}(\vec{P} - \vec{q}_2)\delta(P_u+\omega)|\O\rangle + \text{contact}\,.\notag
\end{align}
We used the kinematics to write
\begin{equation}
    \frac{i}{k_{1v}+P_v+i\epsilon} = \frac{4i\omega}{P^2+s_1+i\epsilon}\,.
\end{equation}
Taking the double discontinuity of \eqref{gaad} gives \eqref{dDiscG}. The contact terms drop out because they can have a single discontinuity (from partial contact terms) but no double discontinuity.

\printbibliography[heading=bibliographyintoc]

\end{document}